\documentclass[letterpaper,10pt,conference]{ieeeconf}
\IEEEoverridecommandlockouts
\usepackage[english]{babel}
\usepackage{amsmath,amssymb,latexsym,amsfonts,mathptmx}
\usepackage{xcolor,graphicx,subfigure}
\usepackage{enumerate}
\usepackage{epsfig}
\usepackage[font = footnotesize]{caption}
\usepackage[lined,linesnumbered,algoruled]{algorithm2e}
\catcode`\<=\active \def<{
     \fontencoding{T1}\selectfont\symbol{60}\fontencoding{\encodingdefault}}
\newcommand{\nin}{\not\in}

\newcommand{\tmtextit}[1]{{\itshape{#1}}}

\newtheorem{theorem}{Theorem}[section]

\newtheorem{assumption}[theorem]{Assumption}

\newtheorem{lemma}[theorem]{Lemma}

\newtheorem{remark}[theorem]{Remark}

\newtheorem{proposition}[theorem]{Proposition}
\newcommand{\oprocendsymbol}{\hbox{$\bullet$}}
\newcommand{\oprocend}{\relax\ifmmode\else\unskip\hfill\fi\oprocendsymbol}

\newcommand{\real}{{\mathbb{R}}}

\renewcommand{\theenumi}{(\roman{enumi})}

\newcommand{\zeros}{\bold{0}}
\newcommand{\ones}{\bold{1}}

\newcommand{\diag}{\operatorname{diag}}

\newcommand{\setdef}[2]{\left\{ #1 \; \big| \; #2\right\}}

\newcommand{\longthmtitle}[1]{\tmtextit{(#1).}}
\newcommand{\myclearpage}{\clearpage}
\renewcommand{\myclearpage}{}

\usepackage[normalem]{ulem}

\renewcommand{\baselinestretch}{.9775}

\allowdisplaybreaks

\myclearpage
\begin{document}

\title{\bf Distributed transient frequency control in power
  networks\thanks{This work is supported by AFOSR Award
    FA9550-15-1-0108.}}

\author{Yifu Zhang and Jorge Cort{\'e}s\thanks{The authors are with
    the Department of Mechanical and Aerospace Engineering, University
    of California, San Diego, CA 92093, USA, {\tt\small
      \{yifuzhang,cortes\}@ucsd.edu}}}
\maketitle

\begin{abstract}
  Modern power networks face increasing challenges in controlling
  their transient frequency behavior at acceptable levels due to low
  inertia and highly-dynamic units. This paper presents a distributed
  control strategy regulated on a subset of buses in a power network
  to maintain their transient frequencies in safe regions while
  preserving asymptotic stability of the overall system. Building on
  Lyapunov stability and set invariance theory, we formulate the
  transient frequency requirement and the asymptotic stability
  requirement as two separate constraints for the control
  input. Hereby, for each bus of interest, we synthesize a controller
  satisfying both constraints simultaneously.  The controller is
  distributed and Lipschitz, guaranteeing the existence and uniqueness
  of the trajectories of the closed-loop system. Simulations on the
  IEEE 39-bus power network illustrate the results.
\end{abstract}

% \begin{keywords}
%   Power systems stability, transient frequency, distributed control,
%   distributed decision making.
% \end{keywords}

\section{Introduction}
To avoid a power system from running underfrequency or to help the
network recover from it, load shedding and curtailment are commonly
employed to balance supply and demand. However, due to inertia, it
takes some time for the energy resources to re-enter a safe frequency
region until the power network eventually converges to steady
state. Hence, during transients, generators are still in danger of
reaching their frequency limits and being tripped, which may in turn
cause blackouts. Therefore, there is a need to analyze the transient
behavior of power networks and design controllers that ensure the safe
evolution of the system.

\subsubsection*{Literature review}
Transient stability refers to the ability of power networks to
maintain synchronism after being subjected to a disturbance, see
e.g.~\cite{PK-JP:04}. The
works~\cite{HDC:11,FD-MC-FB:13,PJM-JH-JK-HJS:14} provide conditions to
ensure synchronicity and investigate their relationship with the
topology of the power network. However, even if network synchronism
holds, system transient trajectory may enter unsafe regions, e.g.,
transient frequency may violate individual generator's frequency
limits, causing generator failure and leading to
blackouts~\cite{PK:94}. Hence, various techniques have been proposed
to improve transient behavior. These include resource re-dispatch with
transient stability constraints~\cite{AA-EBM:06,TTN-VLN-AK:11};
%thyristor-controlled series capacitor compensation to optimize
%transmission impedance and keep power transfer
%constant~\cite{TG-JP:01};
 the use of power system stabilizers to damp
out low frequency inter-machine oscillations~\cite{MAM-HRP-MA-MJH:14},
and placing virtual inertia in power networks to mitigate transient
effects~\cite{TSB-TL-DJH:15,BKP-SB-FD:17}. While these approaches have
a qualitative effect on transient behavior, they do not offer strict
guarantees as to whether the transient frequency stays within a
specific region. Furthermore, the approach in~\cite{TSB-TL-DJH:15}
requires a priori knowledge of the time evolution of the disturbance
trajectories and an estimation of the transient overshoot. Alternative
approaches rely on the idea of identifying the disturbances that may
cause undesirable transient behaviors using forward and backward
reachability analysis, see e.g.,~\cite{MA:14,YCC-ADD:12}
%HC-PJS-SVD:16
and our previous work~\cite{YZ-JC:17-acc}.  The general lack of works
that provide tools for transient frequency control motivates us here
to design feedback controllers for the generators that guarantee
simultaneously the stability of the power network and the desired
transient frequency behavior. Our design is inspired by the
controller-design approach to safety-constrained systems taken
in~\cite{ADA-XX-JWG-PT:16}, where the safety region is encoded as the
zero-sublevel set of a barrier function and safety is ensured by
constraining the evolution of the function along the system
trajectories.

% Most of the state of the art focuses on the design of load shedding
% and curtailment schemes to reach a new stable steady state but
% offers no guarantees on the transient frequency behavior, see
% e.g.~\cite{HY-VV-ZY:03,EM-CZ-SL:15-tac}.  This work focuses on the
% design of feedback controllers for the generators that guarantee
% simultaneously the stability of the power network and a desired
% transient frequency behavior.

\subsubsection*{Statement of contributions}
The main result of the paper is the synthesis of a distributed
controller, available at specific individual generator nodes, that
satisfies the following requirements (i) the closed-loop power network
is asymptotically stable; (ii) for each generator node, if its initial
frequency belongs to a desired safe frequency region, then its
frequency trajectory stays in it for all subsequent time; and (iii)
if, instead, its initial frequency does not belong to the safe region,
then the frequency trajectory enters it in finite time, and once
there, it never leaves.  Our technical approach to achieve this
combines Lyapunov stability and set invariance theory. We first show
that requirement (iii) automatically holds if (i) and (ii) hold true,
and we thereby focus our attention on the latter. For each one of
these requirements, we provide equivalent mathematical formulations
that are amenable to control design. Regarding (i), we consider an
energy function for the power system and formalize (i) as identifying
a controller that guarantees that the time evolution of this energy
function along every trajectory of the dynamics is
non-decreasing. Regarding (ii), we show that this condition is
equivalent to having the controller make the safe frequency interval
forward invariant.  % We then show that the set of controllers
% satisfying (i) and (ii) is nonempty.
Our final step is to use the identified constraints to synthesize a
specific controller that satisfies both and is distributed. The latter
is a consequence of the fact that, for each bus, the constraints only
involve the state of the bus and that of neighboring states.  We show
that the controller is Lipschitz, so that the trajectories of the
closed-loop system exist and are unique for each initial condition,
and that is robust to measurement error and parameter uncertainty.  We
illustrate the performance and design trade-offs of the controller on
the IEEE 39-bus power network. For reasons of space, all proofs are
omitted and will appear elsewhere.

\section{Preliminaries}\label{section:prelimiaries}
In this section we introduce basic notation used throughout the paper
and notions from set invariance and graph theory.

\subsubsection{Notation}%
Let $\real$ and $\real_{\geqslant}$ denote the set of real and
nonnegative real numbers, respectively.  Variables are assumed to
belong to the Euclidean space if not specified otherwise. We let
$\|\cdot\|_{2}$ denote the 2-norm on $\real^{n}$. For a point
$x\in\real^{n}$ and $r>0$, denote
$B_{r}(x)\triangleq\setdef{x'\in\real^{n}}{\|x'-x\|_{2}\leqslant r}$.
Denote $\ones_n$ and $\zeros_n$ in $\real^n$ as the vector of all ones
and zeros, respectively.  For $A\in\mathbb{R}^{m\times n}$, let
$[A]_i$ and $[A]_{i,j}$ denote its $i$th row and $(i,j)$th
element. Given a set $\mathcal{C} \subset \real^{n}$,
$\partial\mathcal{C}$ denotes its boundary. A continuous function
$\alpha:\real\rightarrow \real$ belongs to the 
class-$\mathcal{K}$ functions
%~\cite{ADA-XX-JWG-PT:16} 
if it is
strictly increasing and $\alpha(0)=0$.  Given a differentiable
function $l:\real^{n}\rightarrow\real$, we let $\nabla l$ denote its
gradient. A function $f:\real_{\geqslant
}\times\real^{n}\rightarrow\real^{n},\ (t,x)\rightarrow f(t,x)$ is
Lipschitz in $x$ (uniformly in $t$) if for every $x_{0}\in\real^{n}$,
there exist $L,r>0$ such that $\|f(t,x)-f(t,y)\|_{2}\leqslant
L\|x-y\|_{2}$ for any $x,y\in B_{r}(x_{0})$ and any $t\geqslant 0$.

\subsubsection{Set invariance}
We introduce here notions of forward invariance.  Consider a
nonlinear, non-autonomous system
\begin{align}\label{eqn:nonlinear}
  \dot x=f(t,x), \quad x(0)=x_{0},
\end{align}
where $f:\real_{\geqslant}\times\real^{n}\rightarrow\real^{n}$. We
assume $f$ is piecewise continuous in $t$ and Lipschitz in $x$, so
that the solution of~(\ref{eqn:nonlinear}) exists and is unique.  A
set $\mathcal{C}\in\real^{n}$ is \textit{(forward) invariant} for
system~\eqref{eqn:nonlinear} if for every initial condition $x_{0}\in
\mathcal{C}$, the solution starting from $x_0$ satisfies $x(t)\in
\mathcal{C}$ for all $t\geqslant 0$.  We are particularly interested
in the case where the set $\mathcal{C}$ can be characterized as a
sublevel set of a function.  Suppose $l:\real^{n}\rightarrow\real$ is
continuously differentiable, and let $  \mathcal{C}\triangleq\setdef{x}{l(x)\leqslant 0}.$ The following result states a sufficient and necessary condition for
$\mathcal{C}$ to be forward invariant for~\eqref{eqn:nonlinear}.

\begin{lemma}\longthmtitle{Nagumo's
    Theorem~\cite{FB-SM:08}}\label{lemma:Nagumo}
  Suppose that for all $x\in\mathcal{C}$, there exists $z\in\real^{n}$
  such that $l(x)+\nabla l(x)^{T}z<0$. Furthermore, suppose there
  exists a Lipschitz function
  $\phi:\real^{n}\rightarrow\real^{n}$ such that $\nabla
  l(x)^{T}\phi(x)<0$ for all $x\in\partial\mathcal{C}$.  Then
  $\mathcal{C}$ is forward invariant iff $\nabla
  l(x)^{T}f(t,x)\leqslant 0$ for all $x\in\partial\mathcal{C}$.
\end{lemma}

%The assumptions in Nagumo's Theorem ensure that the set $\mathcal{C}$
%is regular enough to have a well-defined interior and boundary.
% Specifically, the existence of $z$ corresponds to constraint
% qualification, and such an assumption implies that the interior of
% the set $\mathcal{C}$ is simply $\setdef{x}{l(x)< 0}$. The second
% constraint requires the existence of a vector function pointing
% inside $\mathcal{C}$ from each boundary point.

\subsubsection{Graph theory}
We present basic notions in algebraic graph theory
from~\cite{FB-JC-SM:08cor,NB:94}. An undirected graph is a pair
$\mathcal{G} = \mathcal(\mathcal{I},\mathcal{E})$, where $\mathcal{I}
= \{1,\dots,n\}$ is the vertex set and $\mathcal{E}=\{e_{1},\dots,
e_{m}\} \subseteq \mathcal{I} \times \mathcal{I}$ is the edge set.  A
path is an ordered sequence of vertices such that any pair of
consecutive vertices in the sequence is an edge of the graph. A graph
is connected if there exists a path between any two vertices. Two
nodes are neighbors if there exists an edge linking them. Denote
$\mathcal{N}(i)$ as the set of neighbors of node~$i$. For each edge
$e_{k} \in \mathcal{E}$ with vertices $i,j$, the orientation procedure
consists of choosing either $i$ or $j$ to be the positive end of
$e_{k}$ and the other vertex to be the negative end. The incidence
matrix $D=(d_{k i}) \in \mathbb{R}^{m \times n}$ associated with
$\mathcal{G}$ is defined as $ d_{k i} = 1$ if $i$ is the positive end
of $e_{k}$, $ d_{k i} = -1$ if $i$ is the negative end of $e_{k}$, and
$ d_{k i} = 0$ otherwise.
% \begin{align*}
%   d_{k i} = 
%   \begin{cases}
%     1 & \text{if $i$ is the positive end of $e_{k}$},
%     \\
%     - 1 & \text{if $i$ is the negative end of $e_{k}$},
%     \\
%     0 & \text{otherwise}.
%   \end{cases}
% \end{align*}

\section{Problem statement}\label{section:problem-statement}
In this section we introduce the dynamical model for the power network
and state our control objective.

\subsection{Power network model}
The power network is encoded by a connected undirected graph
$\mathcal{G}=(\mathcal{I},\mathcal{E})$, where
$\mathcal{I}=\{1,2,\cdots,n\}$ is the collection of buses and
$\mathcal{E}=\{e_{1},e_{2},\cdots,e_{m}\}\subseteq\mathcal{I}\times\mathcal{I}$
is the collection of transmission lines.  For each node
$i\in\mathcal{I}$, let $\theta_{i}\in\real$ and $\omega_{i}\in\real$
denote its voltage phase angle and shifted voltage frequency relative
to the nominal frequency,
respectively. % (which typically is 50Hz or 60Hz).
We let $p_{i}\in\real$ denote the mechanical power injection for a
generator node and the consumed power injection for a load node.  We
partition buses into $\mathfrak{C}$ and
$\mathcal{I}\backslash\mathfrak{C}$, depending on whether an external
transient control input is available to regulate the frequency
transient behavior of the corresponding bus.  The linearized power
network dynamics described by states of voltage angles and frequencies
is,
% \begin{subequations}\label{eqn:swing-equations-dynamics}
\begin{align}\label{eqn:swing-equations-dynamics}
  \dot\theta_{i}(t) &\hspace{-0.1cm} =\hspace{-0.cm}\omega_{i}(t), \ \forall
  i\in\mathcal{I},
  \\
 M_{i}\dot\omega_{i}(t) & \hspace{-0.1cm}=\hspace{-0.05cm}
  -E_{i}\omega_{i}(t)\hspace{-0.05cm}- \hspace*{-2ex} \sum_{j\in\mathcal{N}(i)}
  \hspace*{-1.5ex} b_{ij} (\theta_{i}(t)\hspace{-0.05cm}-\hspace{-0.05cm}\theta_{j}(t))
  \hspace{-0.05cm}+u_{i}(t)\hspace{-0.05cm}+ p_{i}(t),\
  \hspace{-0.1cm} \forall i\in\mathfrak{C},\notag
  \\
  M_{i}\dot\omega_{i}(t) & \hspace{-0.1cm}=\hspace{-0.05cm}
  -E_{i}\omega_{i}(t)\hspace{-0.05cm}- \hspace*{-2ex} \sum_{j\in\mathcal{N}(i)}
  \hspace*{-1.5ex} b_{ij} (\theta_{i}(t)\hspace{-0.05cm}-\hspace{-0.05cm}\theta_{j}(t))
  \hspace{-0.1cm}+ p_{i}(t),\ \hspace{-0.1cm}\forall
  i\in\mathcal{I}\backslash\mathfrak{C}, \notag
\end{align}
% \end{subequations}
where $b_{ij}>0$ is the susceptance of the line connecting bus $i$
and~$j$, and $M_{i} \in \real_{\geqslant}$ and
$E_{i} \in \real_{\geqslant}$ are the inertia and damping coefficients
of bus $i \in \mathcal{I}$.  For simplicity, we assume that they are
strictly positive for every $i\in\mathcal{I}$.

The above dynamics can be rewritten in a more compact way as
follows. Let $D\in\real^{m\times n}$ be the incidence matrix
corresponding to an arbitrary orientation of the graph. Define
\begin{align}\label{eqn:state-transformation}
  \lambda \triangleq D\theta \in\real^{m}, \quad p\triangleq
  [p_{1},p_{2},\cdots,p_{n}]^{T}\in\real^{n}
\end{align}
as the voltage angle difference vector, and the collection of all
power injections, respectively (here $\theta \triangleq [\theta_{1},
\theta_{2}, \cdots, \theta_{n}]^{T} \in \real^{n}$).  Let
$Y_{b}\in\real^{m\times m}$ be the diagonal matrix whose $k$th
diagonal item represents the susceptance of the transmission line
$e_{k}$ connecting bus $i$ and $j$, i.e., $[Y_{b}]_{k,k}=b_{ij},$ for
$ k=1,2,\cdots, m$.  We re-write the
dynamics~(\ref{eqn:swing-equations-dynamics}) as
\begin{subequations}\label{eqn:dynamics-2}
  \begin{align}
    \dot \lambda (t) &= D\omega(t),\label{eqn:dynamics-2a}
    \\
    M_{i}\dot\omega_{i}(t) &=
    -E_{i}\omega_{i}(t)-[D^{T}Y_{b}]_{i}\lambda(t)+u_{i}(t)+p_{i}(t),
    \ \forall i\in\mathfrak{C},\label{eqn:dynamics-2b}
    \\
    M_{i}\dot\omega_{i}(t)&=-E_{i}\omega_{i}(t)-[D^{T}Y_{b}]_{i}\lambda(t)+p_{i}(t),
    \ \forall
    i\in\mathcal{I}\backslash\mathfrak{C},\label{eqn:dynamics-2c}
    \\
    \lambda(0)&\in\text{range}(D),\label{eqn:initial-condition}
  \end{align}
\end{subequations}
where the initial condition constraint~(\ref{eqn:initial-condition})
is enforced by the
transformation~(\ref{eqn:state-transformation}). When convenient, for
conciseness, we use $ x\triangleq
\left(\lambda,\omega\right)\in\real^{m+n}%,\
% p(t)=[p_{1}(t),p_{2}(t),\cdots,p_{n}(t)]^{T}\in\real^{n}
$ to denote the collection of all states. % and all power injections,
% respectively.
To simply the analysis of stability and convergence of the
dynamics~(\ref{eqn:dynamics-2}) without the controller $u_{i}(t)$, it
is commonly assumed~\cite{CZ-UT-NL-SL:14} that the power
%ARB-VV:00
injection $p_{i}$ is time-invariant.  In order to further investigate
the transient behavior of the dynamics, we relax the time-invariant
condition to the following finite-time convergence condition.

\begin{assumption}\longthmtitle{Finite-time convergence of power
    injections}\label{assumption:finite-time-injection}
  For every $i\in\mathcal{I}$, $p_{i}$ is piecewise continuous and
  becomes constant after a finite time, i.e., $\exists\ \bar t>0$ such
  that $p_{i}(t)=p_{i}^{*}$, for all $i\in\mathcal{I}$ and $t\geqslant
  \bar t$.
\end{assumption}
% Generally speaking, $p_{i}$'s from the generator side are determined
% at %a higher decision layer to specify setpoints for individual
% generator buses %in order to balance the power consumptions given
% from the load side, and %at the same time, to minimize the economic
% cost.

This assumption is motivated by scenarios where the load consumption
changes, causing generators to correspondingly adjust their injections
to balance power consumption, while at the same time minimizing
economic cost.
% In addition, due to the rotor's inertia and unpredictability of some
% generators using renewable energy resources (e.g., wind energy), it
% takes time for generators to reach the targeted setpoints.

Since the dynamics~\eqref{eqn:dynamics-2} is linear, the above
assumption does not affect the convergence result built
on constant power injections for the open loop
system~\eqref{eqn:dynamics-2} without $u_{i}(t)$. Formally, under
Assumption~\ref{assumption:finite-time-injection}, if
$u_{i}\equiv0$ for every $ i\in\mathcal{I}$, then 
$\left(\lambda(t),\ \omega(t)\right)$ converges~\cite{EM:14} to the
unique equilibrium $(\lambda_{\infty},\omega_{\infty}\ones_{n})$, where
$\omega_{\infty}=\frac{\sum_{i=1}^{n}
      p_{i}^{*}}{\sum_{i=1}^{n}E_{i}}$ and $\lambda_{\infty}\in\real^{m}$  is
uniquely determined by $D^{T}Y_{b}\lambda_{\infty}=p^{*}-E\omega_{\infty}\ones_{n}$ and $\   \lambda_{\infty}\in\text{range}(D),$
%  \begin{align}\label{eqn:equilibrium}
 %   D^{T}Y_{b}\lambda_{\infty}=p^{*}-E\omega_{\infty}\ones_{n},\   \lambda_{\infty}\in\text{range}(D),
 % \end{align}
where $p^{*} \triangleq [p_{1}^*, \cdots, p_{n}^*]^{T} \in \real^{n}$,
and $E\triangleq\diag(E_{1},E_{2},\cdots,E_{n})\in\real^{n\times n}$.
% \begin{lemma}\longthmtitle{Stability and convergence of power networks without
% transient controller} \label{lemma:stability-convergence-no-control}
% Under Assumption~\ref{assumption:finite-time-injection}, if
% $u_{i}\equiv0, \ \forall i\in\mathcal{I}$, then the power network
% dynamics~(\ref{eqn:dynamics-2}) is globally stable. Furthermore,
% $\left(\lambda(t),\ \omega(t)\right)$ converge to the unique
% equilibrium $(\lambda_{\infty},\omega_{\infty}\ones_{n})$ where
% $\lambda_{\infty}\in\real^{m}$ and $\omega_{\infty}\in\real$ are
% uniquely determined by
% \begin{subequations}\label{eqn:equilibrium}
% \begin{align}
% D^{T}Y_{b}\lambda_{\infty}&=p^{*}-E\omega_{\infty}\ones_{n},
% \\
% \lambda_{\infty}&\in\text{range}(D),
% \\
% \omega_{\infty}&=\frac{\sum_{i=1}^{n} p_{i}^{*}}{\sum_{i=1}^{n}E_{i}},
% \end{align}
% \end{subequations}
% where $p^{*}\in\real^{n}$ is the collection of $p_{i}^{*}$, and
% $E\triangleq\diag(E_{1},E_{2},\cdots,E_{n})\in\real^{n\times n}$.
% 
%\end{lemma}
\subsection{Control goal}
Our control goal is to design a state-feedback controller for each bus
$i\in\mathfrak{C}$ that preserves the stability and convergence
properties of the system~(\ref{eqn:dynamics-2}) when no external input
$u_{i}$ is present, and at the same time guarantees that the frequency
transient behavior stays within desired safety bounds. We state these
requirements explicitly next.

\subsubsection{Stability and convergence requirement} 
Since the system~\eqref{eqn:dynamics-2} without $u_{i}$ is globally
asymptotically stable (provided $p(t)$ is constant) with respect to
the equilibrium $(\lambda_{\infty},\omega_{\infty}\ones_{n})$, we
require that the system with the proposed controller $u_{i}$ is also
globally asymptotically stable for the same equilibrium, meaning that
$u_{i}$ only affects the transient behavior.

\subsubsection{Invariance requirement}
For each $i\in\mathfrak{C}$, let $\underline\omega_{i},\bar\omega_{i}\in\real$  be lower and upper safe frequency bounds,
where $\underline\omega_{i}<\bar\omega_{i}$.  We require that the
frequency $\omega_{i}(t)$ stays inside the safe region
$[\underline\omega_{i},\bar\omega_{i}]$ for any $t> 0$,
provided that the initial frequency $\omega_{i}(0)$ lies inside
$[\underline\omega_{i},\bar\omega_{i}]$.  This  invariance
requirement corresponds to underfrequency/overfrequency avoidance.
% i.e., if the bus frequencies initially behave normally, then they
% never go beyond the safe region for anytime forward.

\subsubsection{Attractivity requirement}
If, for some $i\in\mathfrak{C}$, the initial frequency
$\omega_{i}(0)\notin[\underline\omega_{i},\bar\omega_{i}]$, then after
a finite time, $\omega_{i}$ enters the safe region and never leaves
afterwards. This requirement corresponds to
underfrequency/overfrequency recovery, i.e., if
underfrequency/overfrequency has already happened, then the controller
should be able to drag the frequency back to the safe region so that,
once it comes back, it never leaves.

The attractivity requirement is automatically satisfied once the
controller meets the first two requirements, provided that
$\omega_{\infty}\in (\underline\omega_{i},\bar\omega_{i})$.  Our goal
is to design a controller that satisfies the above three requirements
and is distributed, in the sense that each bus can implement it using
its own information and that of its neighboring buses and transmission
lines.

% For system~(\ref{eqn:dynamics-2}),our goal is to design a
% distributed controller $u_{i}(t)$ for each $i\in\mathfrak{C}$ such
% that a) the closed-loop system possesses a unique solution; b) the
% system is asymptotically stable; c) for each $i\in\mathfrak{C}$,
% both $\bar{\mathcal{C}}_{i}$ and $\underline{\mathcal{C}}_{i}$ are
% invariant so that their intersection is also invariant, i.e., if
% $\omega_{i}(0)\in [\underline\omega_{i} \bar\omega_{i}]$, then
% $\omega_{i}(t)\in[\underline\omega_{i} \bar\omega_{i}],\ \forall
% t\geqslant 0$;
% d) if for some $i\in\mathfrak{C}$,
% $\omega_{i}(0)\notin[\underline\omega_{i}\bar\omega_{i}]$, then
% there exists a finite time $t_{0}$ such that
% $\omega_{i}(t)\in[\underline\omega_{i} \bar\omega_{i}],\ \forall
% t\geqslant t_0$.

\section{Constraints on controller design}
In this section, we identify constraints on the design of the
controller that provide sufficient conditions to ensure, on  one
hand, the stability and convergence requirement and, on the other
hand, the invariance requirement stated in
Section~\ref{section:problem-statement}.

\subsection{Constraint ensuring stability and convergence}
We establish a stability constraint by identifying an energy function
and restricting the input so that its evolution along every trajectory
of the closed-loop dynamics is monotonically nonincreasing. Moreover,
we show that the convergence holds automatically using the LaSalle
Invariance Principle. The following result states this.
%~\cite{HKK:02}. 

\begin{lemma}\longthmtitle{Sufficient condition for stability and
    convergence}\label{lemma:sufficient-stability-convergence}
  Under Assumption~\ref{assumption:finite-time-injection}, further
  suppose that for every $i\in\mathfrak{C}$,
  $u_{i}:\real^{m+n}\times\real^{n}\rightarrow\real,\ (x,y)\mapsto
  u_{i}(x,y)$ is Lipschitz in $x$ and continuous in $y$.  If for each
  $i\in\mathfrak{C}$, $x\in\real^{m+n}$, $p(t)\in\real^{n}$,
  \begin{subequations}\label{ineq:stabilize-constraints-2}
    \begin{align}
      (\omega_{i}-\omega_{\infty})u_{i}(x,p(t))&\leqslant 0, \
      \text{if }
      \omega_{i}\neq\omega_{\infty},\label{ineq:stabilize-constraints-2a}
      \\
      u_{i}(x,p(t))&=0, \ \text{if
      }\omega_{i}=\omega_{\infty},\label{ineq:stabilize-constraints-2b}
    \end{align}
  \end{subequations}
  then the following results hold 
  \begin{enumerate}
  \item \label{item:solution} The solution of the closed-loop
    system~(\ref{eqn:dynamics-2a})-(\ref{eqn:dynamics-2c}) exists and
    is unique for any $t\geqslant 0$.
  \item\label{intem:convergence-stability}
    $(\lambda(t),\omega(t))\rightarrow(\lambda_{\infty},\omega_{\infty}\ones_{n})$
    as $t\rightarrow \infty$ for any $\omega(0)\in\real^{n}$ and
    $\lambda(0) \in\text{range}(D)$. Furthermore, if $p(t)$ is
    time-independent, then the closed-loop system is globally
    asymptotically stable.
  \end{enumerate}
\end{lemma}

Note that Lemma~\ref{lemma:sufficient-stability-convergence} states
that convergence holds for admissible initial conditions~(\ref{eqn:initial-condition}). % However, since there is no
    % need to consider initial conditions
    % violating~(\ref{eqn:initial-condition}), 
With a little abuse of notation, throughout the rest of the paper, we
refer to~(\ref{ineq:stabilize-constraints-2}) as the asymptotic
stability condition for system~(\ref{eqn:dynamics-2}).

\subsection{Constraint ensuring frequency
  invariance}\label{sec:constraint-freq}
We now define the invariant sets we are interested in. For each
$i \in \mathfrak{C}$, let
$\bar l_{i}, \underline l_{i}: \real^n \rightarrow \real$ be defined
by $\bar l_{i}(x) \triangleq\omega_{i}-\bar\omega_{i}$ and
$\underline l_{i}(x)\triangleq-\omega_{i}+\underline\omega_{i}$.  The
invariant sets are the corresponding sublevel sets, i.e.,
  \begin{align}\label{eqn:barrier-function-unsymmetric}
    \bar{\mathcal{C}}_{i}\triangleq\setdef{x}{\bar l_{i}(x)\leqslant
      0},\quad 
    \underline{\mathcal{C}}_{i}\triangleq\setdef{x}{\underline
      l_{i}(x)\leqslant 0}.
  \end{align}
The invariance condition stated below follows from Nagumo's Theorem.

\begin{lemma}\longthmtitle{Sufficient and necessary condition for
    frequency invariance}\label{lemma:frequency-invariance}
  Suppose the solution
  of~(\ref{eqn:dynamics-2a})-(\ref{eqn:dynamics-2c}) exists and is
  unique. Then for any $i\in\mathfrak{C}$, the sets
  $\bar{\mathcal{C}}_{i}$ and $\underline{\mathcal{C}}_{i}$ are
  invariant if and only if for every $x\in\real^{m+n}$ and $p(t)\in\real^{n}$,
  \begin{subequations}\label{ineq:invariance-condition-unsymmetric-1}
    \begin{align}
      u_{i}(x,p(t))-q_{i}(x,t)\leqslant 0,\
      \text{if }
      \omega_{i}=\bar\omega_{i},\label{ineq:invariance-condition-unsymmetric-1a}
      \\
      -u_{i}(x,p(t))+q_{i}(x,t)\leqslant 0,\
      \text{if }\omega_{i}=\underline
      \omega_{i},\label{ineq:invariance-condition-unsymmetric-1b}
    \end{align}
  \end{subequations}
  where $q_{i}(x,t)\triangleq
  E_{i}\omega_{i}+[D^{T}Y_{b}]_{i}\lambda-p_{i}(t)$.
\end{lemma}

The characterization of Lemma~\ref{lemma:frequency-invariance} points
specifically to the value of the input at the boundary of
$\bar{\mathcal{C}}_{i}$ and $\underline{\mathcal{C}}_{i}$. However,
designing the controller to only become active at such points is
undesirable, as the actuator effort would be discontinuous, affecting
the system evolution. A more sensible policy is to have the controller
become active as the system state gets closer to the boundary of these
sets, and do so in a gradual way. This is captured in the following
result.

\begin{lemma}\longthmtitle{Sufficient condition for frequency
    invariance}\label{lemma:sufficent-frequecy-invariance}
  Suppose the solution
  of~\eqref{eqn:dynamics-2} exists and is
  unique. For every $i\in\mathfrak{C}$, let
  $\bar\omega_{i}^{\text{thr}},\
  \underline\omega_{i}^{\text{thr}}\in\real$ satisfy
  $\underline\omega_{i}<\underline\omega_{i}^{\text{thr}} <
  \bar\omega_{i}^{\text{thr}}<\bar\omega_{i}$. Let $\bar\alpha_{i}$
  and $\underline\alpha_{i}$ be  class-$\mathcal{K}$
  functions.  If for every $x\in\real^{m+n}$ and $p(t)\in\real^{n}$,
  \begin{subequations}\label{ineq:invariance-condition-unsymmetric-4}
    \begin{align}
      (\omega_{i}-\bar\omega_{i}^{\text{thr}})(u_{i}(x,p(t))-q_{i}(x,t))\leqslant
      -\bar\alpha_{i}(\bar l_{i}(x)),
      \label{ineq:invariance-condition-unsymmetric-3-a}
      \end{align}
      if $ \bar\omega_{i}^{\text{thr}}<\omega_{i}\leqslant
      \bar\omega_{i}$, and 
      \begin{align}
        (\underline\omega_{i}^{\text{thr}}-\omega_{i})(-u_{i}(x,p(t))+q_{i}(x,t))\leqslant
        -\underline\alpha_{i}(\underline l_{i}(x)),
        \label{ineq:invariance-condition-unsymmetric-3-b}
    \end{align}
  \end{subequations}
  if $ \underline\omega_{i}\leqslant \omega_{i}<
  \underline\omega_{i}^{\text{thr}}$, then $\bar{\mathcal{C}}_{i}$ and
  $\underline{\mathcal{C}}_{i}$ are invariant.
\end{lemma}

The introduction of the class-$\mathcal{K}$ in
Lemma~\ref{lemma:sufficent-frequecy-invariance} enables the design of
controllers that gradually kick in as the margin for satisfying the
requirement gets increasingly small. In fact,
using~\eqref{eqn:dynamics-2}, we can equivalently
write~\eqref{ineq:invariance-condition-unsymmetric-3-a} as
\begin{align}\label{ineq:derivative-class-K}
  M\dot\omega_{i}\leqslant-\bar\alpha_{i}(\bar
  l_{i}(x))/(\omega_{i}-\bar\omega_{i}^{\text{thr}}), \quad \text{if }
  \bar\omega_{i}^{\text{thr}}<\omega_{i}\leqslant \bar\omega_{i}.
\end{align}
Notice that, as $\omega_{i}$ grows from the threshold
$\bar\omega_{i}^{\text{thr}}$ to the safe bound $\bar\omega_{i}$, the
value of $-\bar\alpha_{i}(\bar
l_{i}(x))/(\omega_{i}-\bar\omega_{i}^{\text{thr}})$ monotonically
decreases from $+\infty$ to 0.  Thus, the constraint on
$\dot\omega_{i}$ becomes tighter (while allowing $\dot\omega_{i}$ to
still be positive) as $\omega_{i}$ approaches $\bar\omega_{i}$, and
when $\omega_{i}$ hits $\bar\omega_{i}$, prescribes $\dot\omega_{i}$
to be nonpositive to ensure safety.
% Loosely speaking, using class-$\mathcal{K}$ functions makes the
% controller early-warning, and at the same time still allows
% $\omega_{i}$ to evolve with some flexibility.
It is interesting to point out the trade-offs present in the choice of
class-$\mathcal{K}$ functions.
% A class-$\mathcal{K}$ function with a
%large derivative, for instance, corresponds to a controller design
%that allows the derivative in~\eqref{ineq:derivative-class-K} to be
%significant near the boundary, at the risk of increasing the
%sensitivity to changes in the state. 
We come back to this point later
in Remark~\ref{rmk:linear-class-K} after introducing our controller
design.

\section{Distributed controller design and analysis}

In this section we introduce our design of transient frequency
control. We characterize the continuity properties of the proposed
distributed controller and show that it meets the asymptotic stability
and the frequency invariance conditions.

\subsection{Controller synthesis for transient frequency control}

Our controller design builds on the
conditions~\eqref{ineq:stabilize-constraints-2}
and~\eqref{ineq:invariance-condition-unsymmetric-4}. Our next result
formally introduces it and characterizes its continuity properties.

\begin{proposition}\longthmtitle{Distributed frequency
    controller}\label{prop:L}
  Under Assumption~\ref{assumption:finite-time-injection}, for each
  $i\in\mathfrak{C}$, let
  \begin{align}\label{eqn:stability-transient-controller-Lipschitz-4}
    u_{i}(x,p(t)) = 
    \begin{cases}
      \min\{0,\frac{-\bar\alpha_{i}(\bar
        l_{i}(x))}{\omega_{i}-\bar\omega_{i}^{\text{thr}}}+q_{i}(x,t)\}
      & \omega_{i}>\bar\omega_{i}^{\text{thr}},
      \\
      0 & \underline\omega_{i}^{\text{thr}}\leqslant
      \omega_{i}\leqslant \bar\omega_{i}^{\text{thr}},
      \\
      \max\{0,\frac{\underline\alpha_{i}(\underline
        l_{i}(x))}{\underline\omega_{i}^{\text{thr}}-\omega_{i}}+q_{i}(x,t)\}
      & \omega_{i}<\underline\omega_{i}^{\text{thr}},
    \end{cases}
  \end{align}
  where $\bar\alpha_{i}$ and $\underline\alpha_{i}$ are any Lipschitz
  class-$\mathcal{K}$ functions defined on $\real$. Then $u_{i}$ is
  Lipschitz in  its first argument.
\end{proposition}

Note that the
controller~\eqref{eqn:stability-transient-controller-Lipschitz-4} is
distributed, because it only requires for each controllable bus $i \in
\mathfrak{C}$ to share information with buses it is connected to in
the power network to compute the term $[D^{T}Y_{b}]_{i}\lambda$, which
corresponds to the aggregate power flow injected at node~$i$ from its
neighboring nodes.

\begin{remark}\longthmtitle{Extension to  nonlinear power
    flows}\label{rmk:control-realization} 
  An interesting observations is that, instead of requiring knowledge
  of $\lambda$ and the susceptance matrix $Y_{b}$ to evaluate the
  aggregate power flow injected by neighbors, this value could simply
  be measured by the node itself. This observation also opens the way
  to extend the controller design to nonlinear
  models~\cite{JM-JWB-JRB:08},
%  , FD-MC-FB:13,ST-MB-CDP:16
 where one
  simply replaces the linearized aggregate power flow
  $[D^{T}Y_{b}]_{i}\lambda$ by the corresponding nonlinear power flow,
  ensuring frequency invariance. % In such cases, the use of nonlinear
  % network models~\cite{FD-MC-FB:13,ST-MB-CDP:16} implies that the
  % controller only locally stabilizes the system. 
  \oprocend
\end{remark}

The next result shows that the proposed distributed controller
achieves the objectives identified in
Section~\ref{section:problem-statement} regarding asymptotic stability
and frequency invariance.

\begin{theorem}\longthmtitle{Transient frequency
    control}\label{thm:decentralized-controller}
  Under Assumption~\ref{assumption:finite-time-injection}, let
  $\omega_{\infty}\in(\underline\omega^{\text{thr}}_{i},\bar\omega^{\text{thr}}_{i})$
  and consider the evolution of the system~\eqref{eqn:dynamics-2} with
  the
  controller~\eqref{eqn:stability-transient-controller-Lipschitz-4}.
  Then
  \begin{enumerate}
  \item\label{item:exist-unique} The solution of the 
    system exists and is unique.
  \item\label{item:asy-stability} The state $\left(\lambda(t),\
      \omega(t)\right)$ converges to $(\lambda_{\infty},\
    \omega_{\infty}\ones_{n})$. Furthermore, if $p(t)$ is
    time-independent, then the closed-loop system is globally
    asymptotically stable.
  \item\label{item:finite-time-active} The controllers become inactive
    in finite time, i.e., there exists a time $t_{0}>0$ such that
    $u_{i}(x,p(t))=0$ for all $t\geqslant t_{0}$ and for all
    $i\in\mathfrak{C}.$
  \item\label{item:frequency-invariant} For any $i\in\mathfrak{C}$, if
    $\omega_{i}(0)\in[\underline\omega_{i},\bar\omega_{i}]$, then
    $\omega_{i}(t)\in[\underline\omega_{i},\bar\omega_{i}]$ for all
    $t> 0$.
  \item\label{item:frequency-attraction} For any $i\in\mathfrak{C}$,
    if $\omega_{i}(0)\nin[\underline\omega_{i},\bar\omega_{i}]$, then
    there exists a finite time $t_{1}>0$ such that for all $t>t_{1}$,
    it holds that
    $\omega_{i}(t)\in[\underline\omega_{i},\bar\omega_{i}]$. In
    addition, convergence towards
    $[\underline\omega_{i},\bar\omega_{i}]$ is monotonic.
    % \item $u_{i}$ can be no-zero only when
    %   $\omega_{i}(t)\nin[\underline\omega^{\text{thr}}_{i},\bar\omega^{\text{thr}}_{i}]$.
  \end{enumerate}
  % In addition, if we apply
  % controller~(\ref{eqn:stability-transient-controller-Lipschitz-4})
  % to the nonlinear dynamics~(\ref{eqn:dynamics-2}), then all the
  % above statements still hold, expect that
  % for~\ref{item:asy-stability}, the closed-loop system now is only
  % locally asymptotically stable.
\end{theorem}

Note that from
Theorem~\ref{thm:decentralized-controller}\ref{item:asy-stability},
the controller only affects the transient behavior of the system and,
from
Theorem~\ref{thm:decentralized-controller}\ref{item:finite-time-active},
it becomes inactive in finite time. These properties guarantee that,
with the transient controller in place, the static power injection
$p_i^*$ at every bus is met.

\begin{remark}\longthmtitle{Performance trade-offs in selection of
     class-$\mathcal{K}$ functions}\label{rmk:linear-class-K}
  {\rm As we pointed out in Section~\ref{sec:constraint-freq}, the
    choice of  class-$\mathcal{K}$ functions affects the
    system behavior. To illustrate this, consider the linear choice
    $\bar\alpha_{i}=\underline\alpha_{i}:\real\rightarrow\real,\ s
    \mapsto \gamma_{i}s$, where $\gamma_{i}>0$ is tunable.  Notice
    that a smaller $\gamma_{i}$ leads to more stringent requirements
    on the derivative of the frequency.  To see this, note that
    $u_{i}(x,p(t))$ can be non-zero only when either of the following
    two conditions happen: $\frac{-\bar\alpha_{i}(\bar
        l_{i}(x))}{(\omega_{i}-\bar\omega_{i}^{\text{thr}})}+q_{i}(x,t)<0\text{
        and } \omega_{i}>\bar\omega_{i}^{\text{thr}},$ or, $\frac{\underline\alpha_{i}(\underline
        l_{i}(x))}{\underline\omega_{i}^{\text{thr}}-\omega_{i}}+q_{i}(x,t)>0\text{
        and } \omega_{i}<\underline\omega_{i}^{\text{thr}}.$
    % \begin{subequations}\label{sube:ineq:controller-react}
%     \begin{align*}
%       \frac{-\bar\alpha_{i}(\bar
%         l_{i}(x))}{(\omega_{i}-\bar\omega_{i}^{\text{thr}})}+q_{i}(x,t)<0\text{
%         and } \omega_{i}>\bar\omega_{i}^{\text{thr}},
%       \\
%       \frac{\underline\alpha_{i}(\underline
%         l_{i}(x))}{\underline\omega_{i}^{\text{thr}}-\omega_{i}}+q_{i}(x,t)>0\text{
%         and } \omega_{i}<\underline\omega_{i}^{\text{thr}}.
%     \end{align*}
    % \end{subequations}
    In the first case, the term $\frac{-\bar\alpha_{i}(\bar
      l_{i}(x))}{(\omega_{i}-\bar\omega_{i}^{\text{thr}})} =
    \frac{\gamma_{i}(\bar\omega_{i}-\omega_{i})}{\omega_{i} -
      \bar\omega_{i}^{\text{thr}}}>0$ becomes smaller as $\gamma_{i}$
    decreases, making its addition with $q_{i}(x,t)$ closer to being
    negative, and resulting in an earlier activation of $u_{i}$.  The
    second case follows similarly.  A small $\gamma_{i}$ may also lead
    to high control magnitude because it prescribes a smaller bound on
    the frequency derivative, which in turn may require a larger
    control effort.
    % For instance, when $\omega_{i}>\bar\omega_{i}^{\text{thr}}$,
    % since $u_{i}$ satisfies invariance
    % condition~\eqref{ineq:invariance-condition-unsymmetric-1}, one
    % has~\eqref{ineq:derivative-class-K} holds, which, if
    % $\bar\alpha_{i}$ is linear, is equivalent to
    % \begin{align*}
    %   M\dot\omega_{i}\leqslant\gamma_{i}(\bar\omega_{i} -
    %   \omega_{i})/(\omega_{i}-\bar\omega_{i}^{\text{thr}}),\
    %   \bar\omega_{i}^{\text{thr}}<\omega_{i}\leqslant \bar\omega_{i}.
    % \end{align*}
    % Intuitively, when $\gamma_{i}$ is positive but close to $0$, from the
    % above inequality, although $\dot\omega_{i}$ is still allowed to be
    % positive, its value is tightly constrained, making $\omega_{i}$
    % possibly increase but with a very slow growth rate. Hence, compared
    % with a $u_{i}$ with large $\gamma_{i}$ that allows $\omega_{i}$ grow
    % faster, it takes extra efforts for controller with small $\gamma_{i}$
    % to maintain the slow growth rate, causing large control magnitude.
    However, choosing a large $\gamma_{i}$ may cause the controller to
    be highly sensitive to $\omega_{i}$. This is because the absolute
    value of the partial derivative of $\frac{-\bar\alpha_{i}(\bar
      l_{i}(x))}{(\omega_{i}-\bar\omega_{i}^{\text{thr}})}$
    (resp. $\frac{\underline\alpha_{i}(\underline
      l_{i}(x))}{\underline\omega_{i}^{\text{thr}}-\omega_{i}}$) with
    respect to $\omega_{i}$ grows proportionally with $\gamma_{i}$;
    consequently, when $u_{i}(x,p(t))$ is non-zero, its sensitivity
    against $\omega_{i}$ increases as $\gamma_{i}$ grows, resulting in
    low tolerance against slight changes in~$\omega_{i}$.  } \oprocend
\end{remark}

\subsection{Robustness to measurement and parameter uncertainty}

Here we study the controller performance under measurement and
parameter uncertainty. This is motivated by scenarios where the state
$x$ or the power injection $p_{i}$ may not be precisely measured, or
scenarios where some system parameters, like the damping coefficient,
is only approximately known.  Formally, we let $\hat x$, $\hat p$, and
$\hat E$ be the measured or estimated state, power injection, and
damping parameters, respectively. For every $i\in\mathfrak{C}$, we
introduce the error variables
%\begin{%subequations}%\label{sube:eqn:uncertain-model}
\begin{align*}
  \epsilon^{\omega}_{i}&\triangleq \hat\omega_{i}-\omega_{i},\   &\epsilon^{\lambda}_{i}\triangleq
  [D^{T}Y_{b}]_{i}\hat\lambda-[D^{T}Y_{b}]_{i}\lambda,
  \\
  \epsilon^{p}_{i}&\triangleq \hat p_{i}-p_{i},\   &\epsilon^{E}_{i}\triangleq \hat E_{i}-E_{i}.\hspace{1.94cm}
\end{align*}
%\end{subequations}
We make the following assumption regarding the
error.

\begin{assumption}\longthmtitle{Bounded
    uncertainties}\label{assumption:bounded-uncertain}
  For each $i\in\mathfrak{C}$,
  \begin{enumerate}
  \item the uncertainties can be bounded by
    $|\epsilon^{\omega}_{i}(t)|\leqslant \bar\epsilon^{\omega}_{i}$,
    $|\epsilon^{\lambda}_{i}(t)|\leqslant \bar\epsilon^{\lambda}_{i}$,
    $|\epsilon^{p}_{i}(t)|\leqslant \bar\epsilon^{p}_{i}$, and
    $|\epsilon^{E}_{i}(t)|\leqslant \bar\epsilon^{E}_{i}$ for all
    $t\geqslant 0$,

  \item the synchronized frequency $\omega_{\infty} \in
    [\underline\omega_{i}^{\text{thr}} +
    \bar\epsilon^{\omega}_{i},\bar\omega_{i}^{\text{thr}}-\bar\epsilon^{\omega}_{i}]$, 
  \item $\bar\epsilon^{\omega}_{i}<
    \min\{\bar\omega_{i}-\bar\omega_{i}^{\text{thr}},
    \underline\omega_{i}^{\text{thr}}-\underline\omega_{i}\}$.
  \end{enumerate}
\end{assumption}

For convenience, we use~$\hat u_{i}(\hat x,\hat p(t))$ to refer to the
controller with the same functional expression
as~\eqref{eqn:stability-transient-controller-Lipschitz-4} but
implemented with approximate parameter values and evaluated at the
inaccurate state $\hat x$ and power injection $\hat p$.  The next
result shows that $\hat u_{i}$ still stabilizes the power network and
enforces the satisfaction of a relaxed frequency invariance
condition. For simplicity, we restrict our attention to linear
class-$\mathcal{K}$ functions in the controller design.

\begin{proposition}\longthmtitle{Robust stability and frequency
    invariance under uncertainty}\label{prop:robust-uncertainty}
  Under Assumption~\ref{assumption:bounded-uncertain}, consider the
  evolution of~\eqref{eqn:dynamics-2} under $\hat u_{i}$ for each
  $i\in\mathfrak{C}$. Then,
  \begin{enumerate}
  \item\label{item:robust-stability} The state $\left(\lambda(t),\
      \omega(t)\right)$ converges to $(\lambda_{\infty},\
    \omega_{\infty}\ones_{n})$. Furthermore, if $p(t)$ is
    time-independent, then the closed-loop system is globally
    asymptotically stable.
  \item\label{item:robust-invariance} For each $i\in\mathfrak{C}$, let
    $\bar\alpha_{i}(s) = \underline\alpha_{i}(s)=\gamma_{i}s$.  Then,
    if there exists $\Delta>0$ such that the error bounds satisfy
    \begin{subequations}\label{sube:ineq:robust-invariance}
      \begin{align}
        \hspace{-1.2cm} \frac{-\gamma_{i}(\bar\epsilon^{\omega}_{i} +
          \Delta)}{\bar\omega_{i}-\bar\omega_{i}^{\text{thr}} +
          \Delta+\bar\epsilon^{\omega}_{i}}+\bar\epsilon^{E}_{i}(\Delta+\bar\omega_{i})
        + \hat E_{i}\bar\epsilon^{\omega}_{i}+\bar\epsilon^{\lambda}_{i} +
        \bar\epsilon^{p}_{i}\leqslant
        0,\label{sube:ineq:robust-invariance-a}
        \\
        \hspace{-1.2cm} \frac{-\gamma_{i}(\bar\epsilon^{\omega}_{i} +
          \Delta)}{\underline\omega_{i}^{\text{thr}}-\underline\omega_{i}
          + \Delta +
          \bar\epsilon^{\omega}_{i}}+\bar\epsilon^{E}_{i}(\Delta-\underline\omega_{i})+\hat
        E_{i}\bar\epsilon^{\omega}_{i}+\bar\epsilon^{\lambda}_{i}+\bar\epsilon^{p}_{i}\leqslant
        0,\label{sube:ineq:robust-invariance-b}
      \end{align}
       then $\omega_{i}(t) \in
      [\underline\omega_{i}-\Delta, \bar\omega_{i}+\Delta]$ for all
      $t> 0$, provided $\omega_{i}(0) \in
      [\underline\omega_{i}-\Delta,\bar\omega_{i}+\Delta]$.
    \end{subequations}
  \end{enumerate}
\end{proposition}

% \marginJC{As a general critique of what we do, one could say that we
%   fix the transient stability problem by requiring arbitrarily large
%   control effort -- a hardly surprising result. Can't we deal with
%   with bounded inputs or quantify the control effort or something that
%   addresses this shortcoming?}
% \marginy{I only put the uncertainty part here as the magnitude bound result is too ugly- I cannot express the magnitude explicitly as a function of, say, size the disturbance, but the solution of some opti. problem. Besize, we need to assume constant power injection.}
% \marginy{I found one way to deal with this a few hours ago. We are
%   able to bound the control maginitude in terms of how far the initial
%   state from the e.q. point, as well as how far the power inejection
%   from its nominal value. Currently, I have only established this
%   result on the linear case, assuming that the power injection is
%   constant. Extending this result to time-varying power injection is
%   easy, but extending it to time-varying injection with nonlinear
%   dyanmics is tough.}

\section{Simulations}

\begin{figure}[htb]
  \centering%
  \vspace{-0.3cm}
  \includegraphics[width=0.9\linewidth]{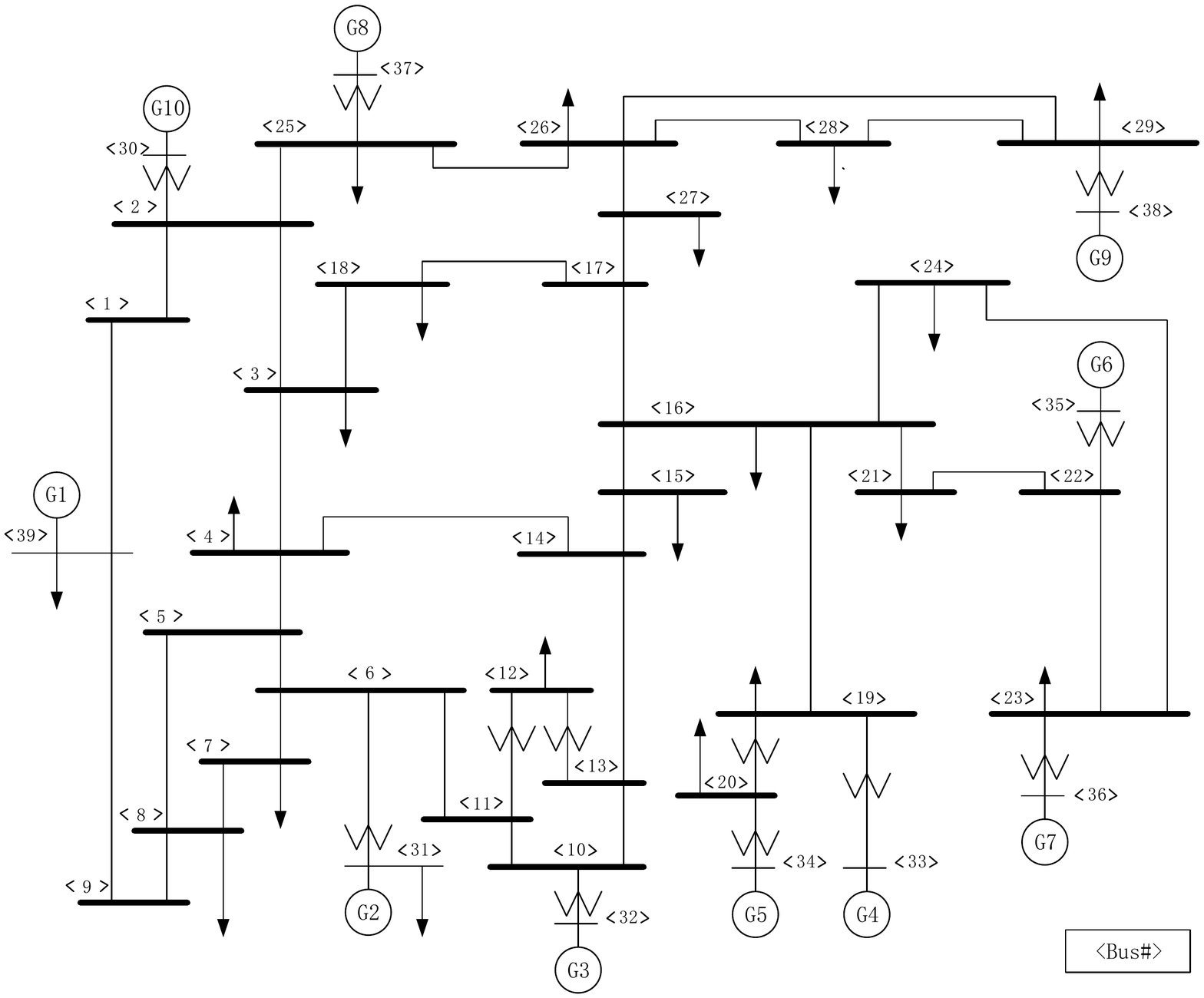}
    \vspace{-0.3cm}
  \caption{IEEE 39-bus power network.}\label{fig:IEEE39bus}
\vspace*{-2ex}
\end{figure}

We illustrate the performance of our control design in the IEEE 39-bus
power network displayed in Figure~\ref{fig:IEEE39bus}.  The network
consists of 46 transmission lines and 10 generators, serving a load of
approximately 6GW.  Instead of the linear
model~\eqref{eqn:dynamics-2}, in the simulation we employ the
nonlinear swing equations~\cite{JM-JWB-JRB:08} to illustrate the
suitability of the proposed controller to even more general scenarios
(cf.  Remark~\ref{rmk:control-realization}).  We take the values of
susceptance $b_{ij}$ and rotational inertia~$M_{i}$ for generator
nodes from the Power System Toolbox~\cite{KWC-JC-GR:09}. We also use
this toolbox to assign the initial power injection $p_{i}(0)$ for
every bus.  We assign all non-generator buses an uniform small inertia
$M_{i}=0.1$. We let the damping parameter to be $E_{i}=1$ for all
buses.  The initial state $(\lambda (0),\omega(0))$ is chosen to be
the equilibrium with respect to the initial power injections.
%  Let
%$\mathfrak{C}=\{30,31,32,33\}$ be the collection of indexes of generators %with proposed transient controller. 
We implement on each generator whose index in
$\mathfrak{C}=\{30,31,32,33\}$ the distributed controller defined
in~(\ref{eqn:stability-transient-controller-Lipschitz-4}) to tune its
transient frequency behavior. The controller parameters are set up as
follows: for every $i\in\mathfrak{C}$, we let
$\bar\alpha_{i}(s) = \underline\alpha_{i}(s)=\gamma_{i}s$, with
$\gamma_{i}=2$, $\bar\omega_{i}=-\underline\omega_{i}=0.2$Hz and
$\bar\omega_{i}^{\text{thr}}=-\underline\omega_{i}^{\text{thr}}=0.1$Hz.
The nominal frequency is 60Hz, and hence the safe frequency region is
$[59.8\text{Hz},\ 60.2\text{Hz}]$.

\begin{figure*}[tbh!]
  \centering
  \subfigure[\label{frequency-response-no-control-generator}]{\includegraphics[width=.22\linewidth]{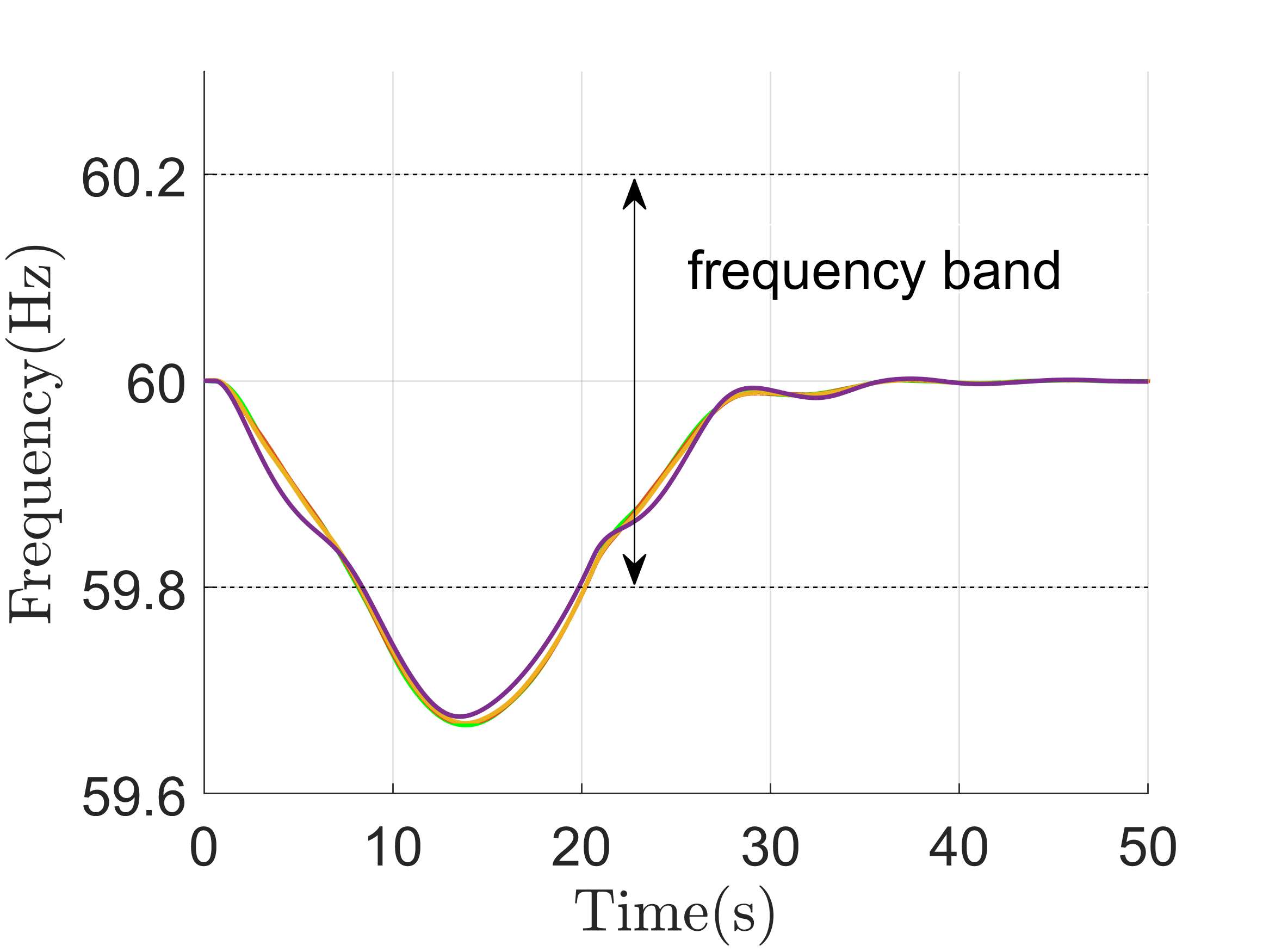}}
  \subfigure[\label{frequency-response-with-control-generator}]{\includegraphics[width=.22\linewidth]{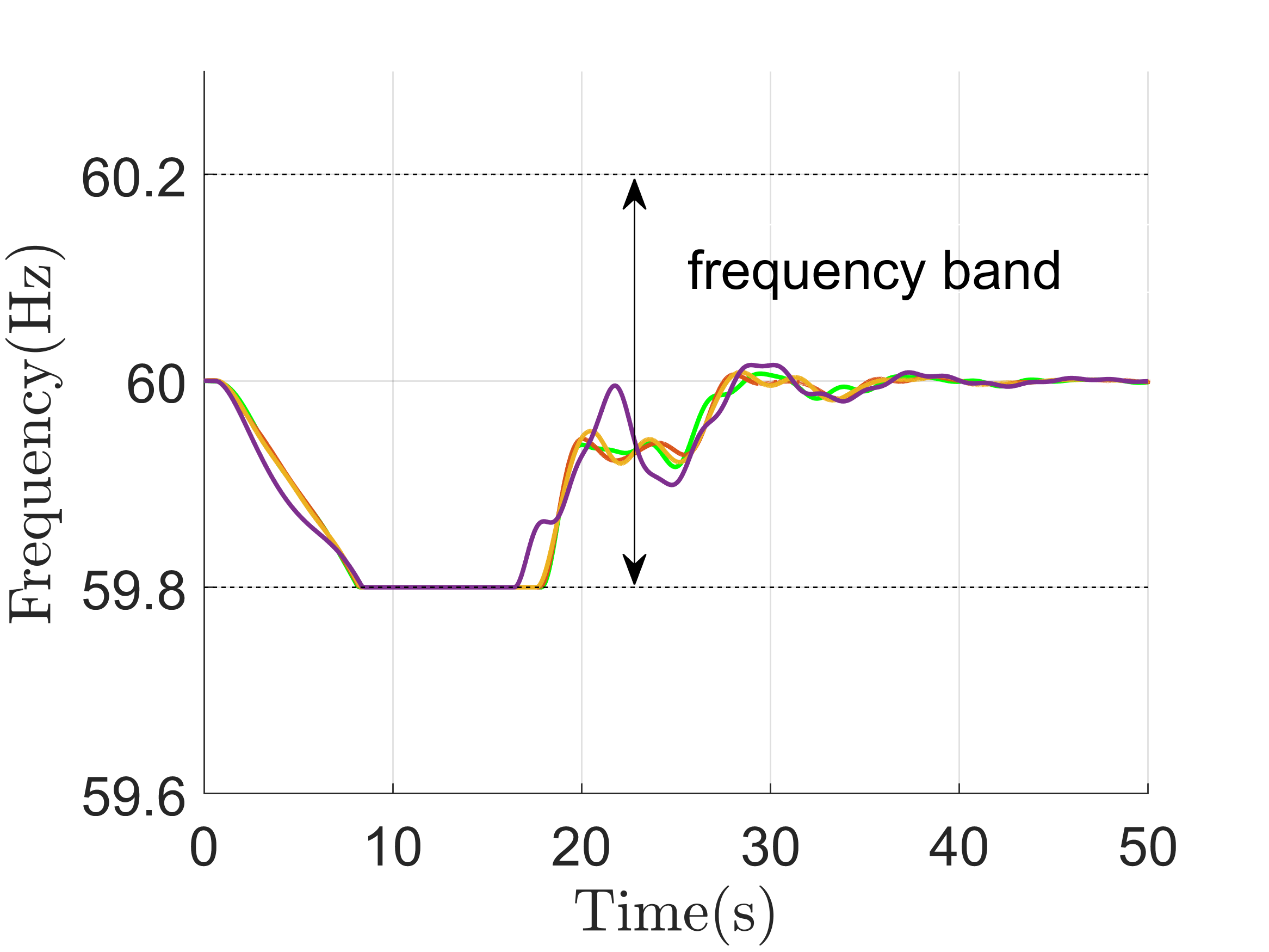}}
  \subfigure[\label{input-trajectories}]{\includegraphics[width=.22\linewidth]{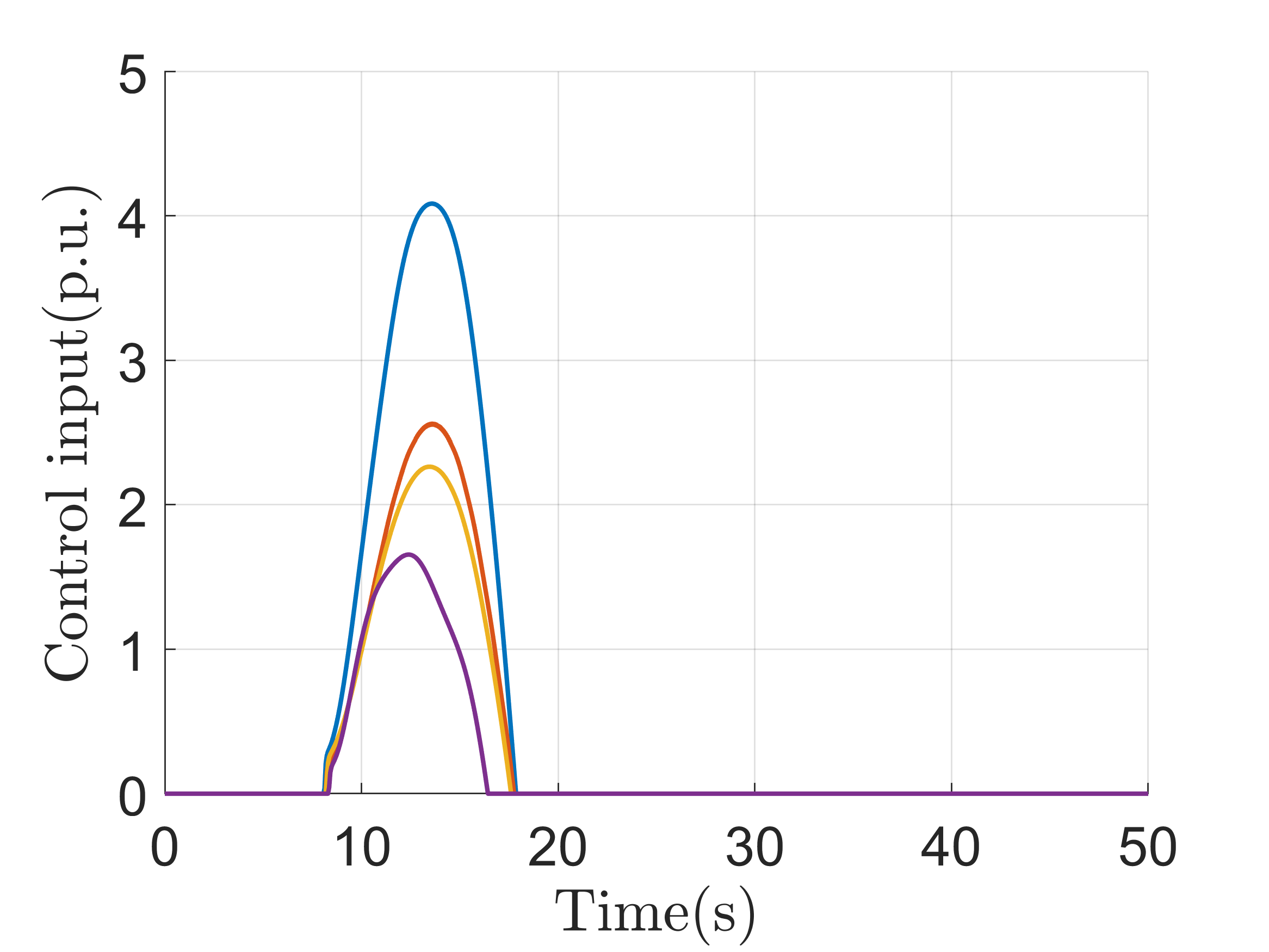}}
  \subfigure[\label{input-trajectories-robust}]{\includegraphics[width=.22\linewidth]{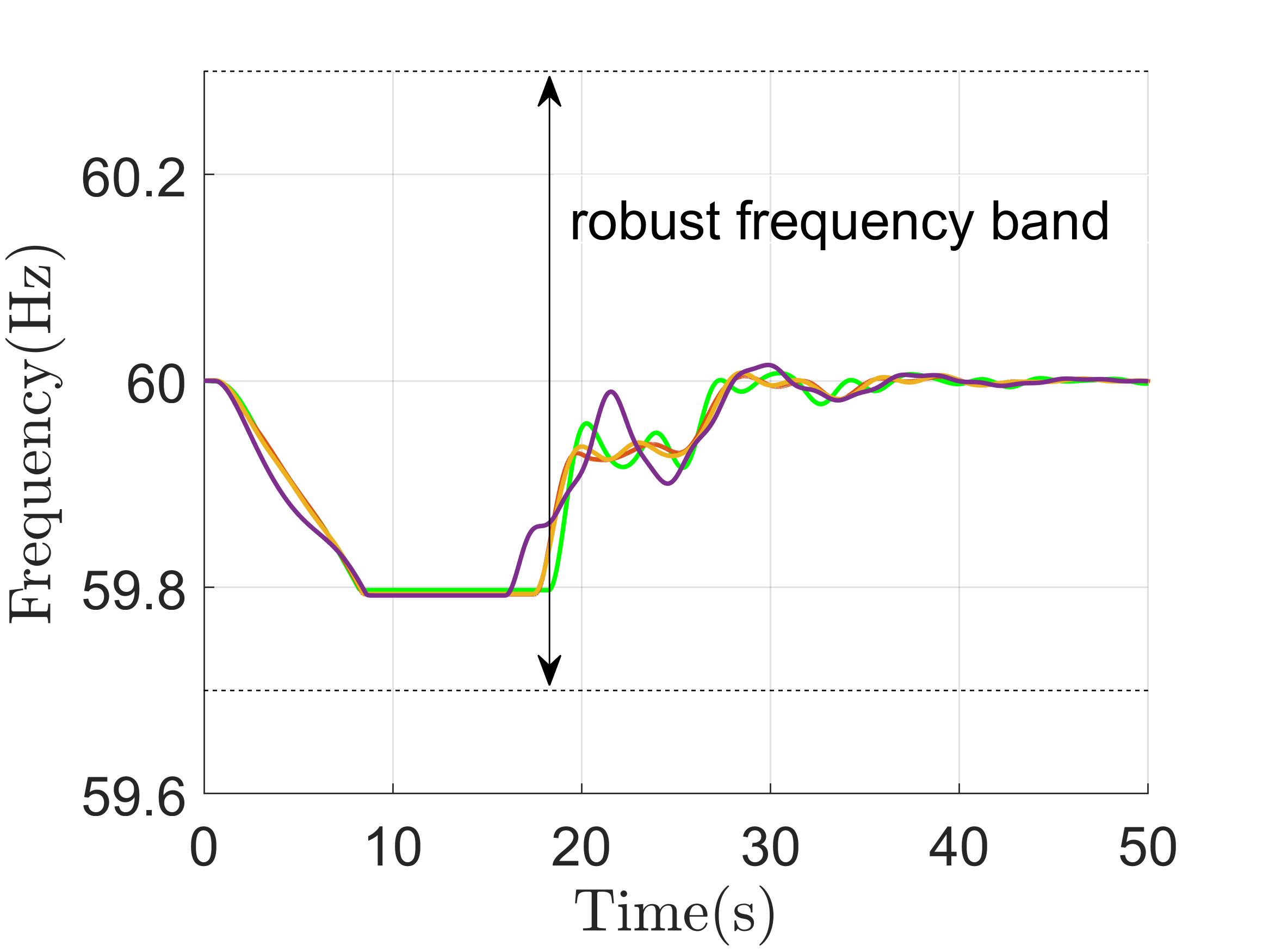}}
  \vspace{-0.3cm}
  \caption{Frequency and control input trajectories with and without
    transient controller. 
    %Plot~\subref{frequency-response-no-control-generator}
    %shows the frequency trajectories of the 4 generators, indexed from
    %30 to 33, without the transient controller~\eqref{eqn:stability-transient-controller-Lipschitz-4}, %with all of them going
    %beyond the lower safe frequency bound. With the transient
    %controller,
    %plot~\subref{frequency-response-with-control-generator} shows that
    %all frequency trajectories stay within the safe
    %bound. Plot~\subref{input-trajectories} shows the corresponding
    %trajectories of the control
    %inputs. Plot~\subref{input-trajectories-robust} shows the
    %controller performance under parameter uncertainty and errors in
    %the power injection approximation.
    }\label{fig:trajectories}
\vspace*{-1ex}
\end{figure*}

We first show that the proposed controller is able to maintain the
targeted generator frequencies within the safe region, provided that
these frequencies are initially in the safe region. We perturb all
non-generator nodes by a sinusoidal power injection whose magnitude is
proportional to the corresponding node's initial power
injection. Specifically, for every $i\in\{1,2,\cdots,29\}$,
\vspace{-0.1cm}
\begin{align*}%\label{eqn:disturbance}
  p_{i}(t)=
  \begin{cases}
    p_{i}(0) & \hspace{-0.7cm}\text{if $t\leqslant0.5$ or $t\geqslant
      20.5$,}
    \\
    \left(1+0.3\sin(\frac{\pi}{20} (t-0.5))\right)p_{i}(0) &
    \text{otherwise.}
  \end{cases}
\end{align*}
For $i\in\{30,31,\cdots,39\}$, $p_{i}(t)$ remains constant all the
time.
Figure~\ref{fig:trajectories}\subref{frequency-response-no-control-generator}
shows the frequency responses of the 4 generators without the
transient controller. One can easily see that all trajectories exceed
the 59.8Hz lower frequency bound. For comparison,
Figure~\ref{fig:trajectories}\subref{frequency-response-with-control-generator}
shows the trajectories with the transient
controller~(\ref{eqn:stability-transient-controller-Lipschitz-4}),
where one can see that all remain within the safe frequency region.
Figure~\ref{fig:trajectories}\subref{input-trajectories} displays the
corresponding input trajectories, which converge to 0 in finite time,
as stated in
Theorem~\ref{thm:decentralized-controller}(iii). % Figure~\ref{fig:trajectories}\subref{frequency-response-with-control-all-nodes}
% are the trajectories of all 39 nodes, which converge to the nominal
% frequency 60Hz as the power injections are eventually re-balanced.
The control input with highest overshoot corresponds to the
generator~$33$.  Its large magnitude, compared to the other inputs, is
due to
% the low inertia $M_{33}$, making $\omega_{33}$ responde rapidly
% against %disturbances, and indirectly having
the power flow in the adjacent edge $(19,33)$ evolving relatively far
from the nominal value, resulting in a large power injection
fluctuation to node $33$, which further causes a large control effort
for compensation.  We also illustrate the robustness of the controller
against uncertainty. We have each controller employ $\hat E_{i}=2$ and
$\hat p_{i}(t)=1.1p_{i}(t)$, corresponding to $100\%$ and $10\%$
deviations on droop coefficients and power injections, respectively.
%
% \marginJC{No state uncertainty? I bet you that with state uncertainty,
%   we don't get perfect convergence, as displayed in the figure. Yet
%   the proposition above says that we do??}
% %
% \marginy{There is no measurement uncertainty from state in this
%   simulation section. Frequencies don't converge to 60Hz in Fig.2 (d),
%   which can be barely seen, since in this set-up, power supply no more
%   equals power consumption even time goes to infinite, and hence the
%   synchronized frequency is no more 60Hz.If the consumption and supply
%   are balanced within finite time, and there is some state
%   uncertainty, frequencies should exactly converge to 60Hz. See
%   colored proof in the proposition.}
%
Figure~\ref{fig:trajectories}\subref{input-trajectories-robust}
illustrates the frequency trajectories of the 4 controlled generators.
Since condition~\eqref{sube:ineq:robust-invariance} is satisfied with
$\Delta=0.1$Hz, Proposition~\ref{prop:robust-uncertainty} ensures that
the frequency interval with invariance guarantee is now % relaxed from
 % $[59.8\text{Hz},60.2\text{Hz}]$ to
$[59.7\text{Hz},60.3\text{Hz}]$ (in fact, the four trajectories only
slightly exceed $59.8$Hz).

\begin{figure}[tbh!]
\vspace{-0.1cm}
  \centering
  \subfigure[\label{IEEE39-omega-trajectories-vs-gamma}]{\includegraphics[width=.45\linewidth]{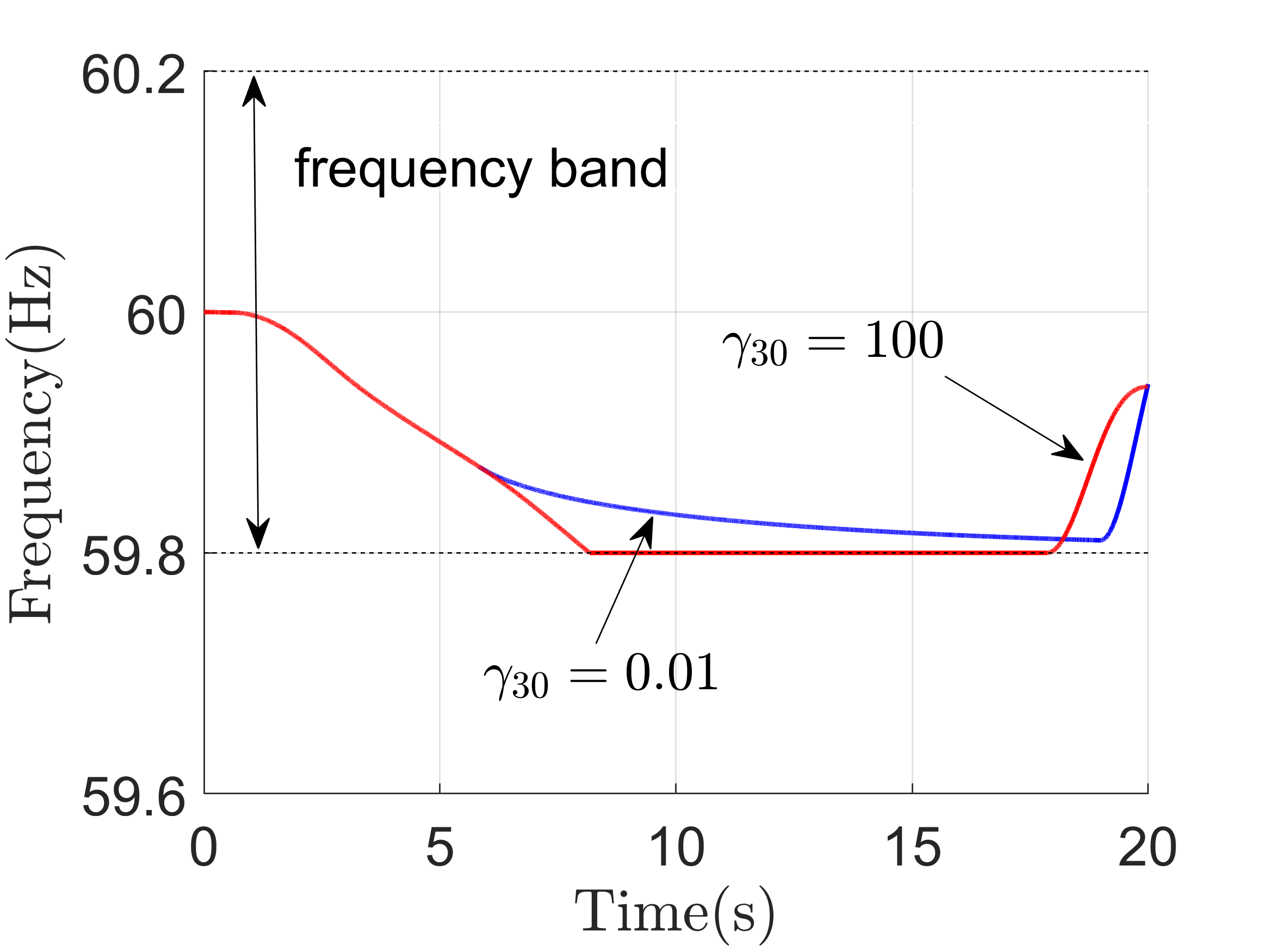}}
  \subfigure[\label{IEEE39-input-trajectories-vs-gamma}]{\includegraphics[width=.45\linewidth]{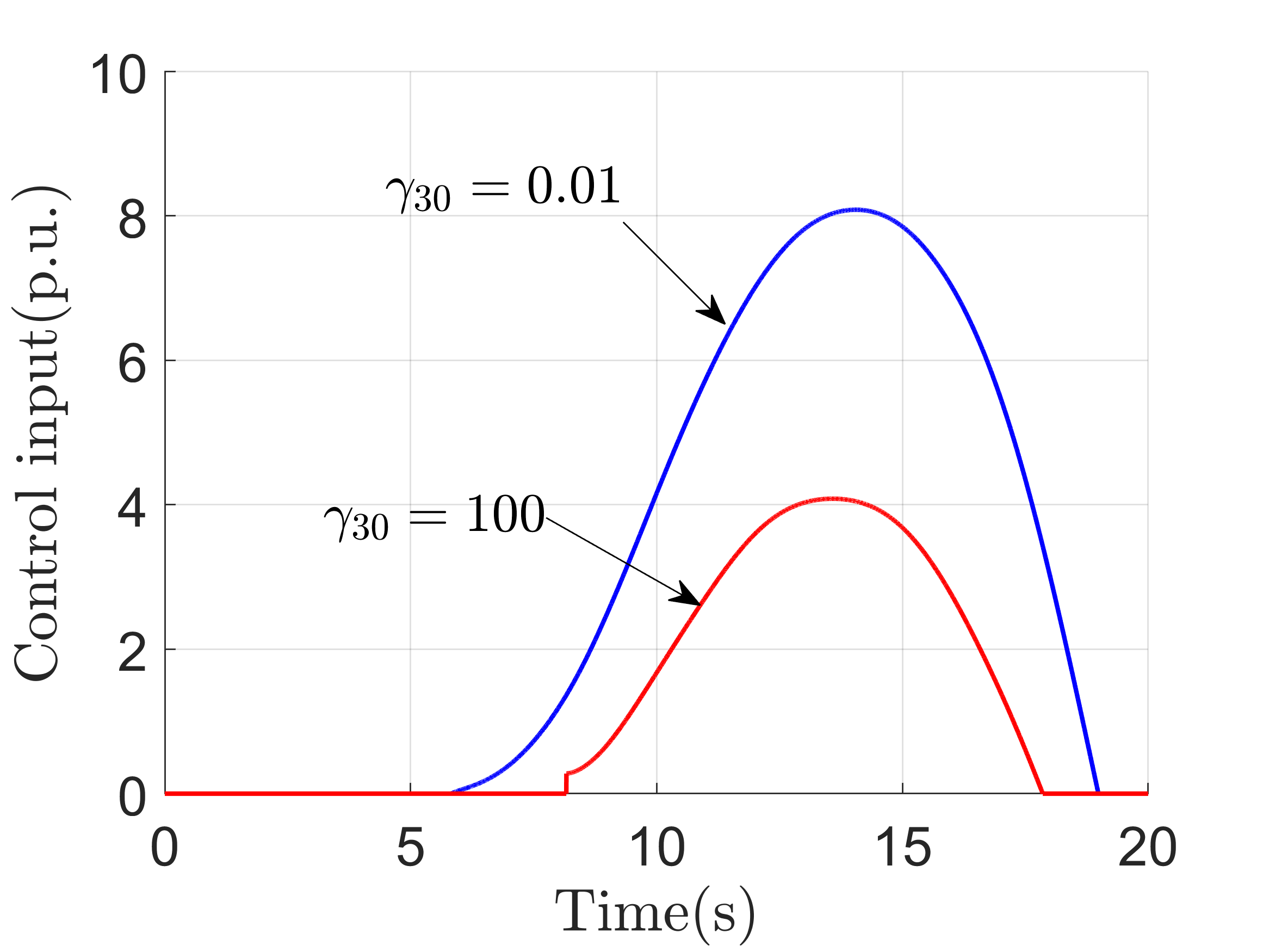}}
    \vspace{-0.3cm}
  \caption{Frequency and control input trajectories at node $30$ with
    linear class-$\mathcal{K}$ function with slope $\gamma_{30}=0.01$ and $10$, respectively.  
%We observe from
    %plot~\subref{IEEE39-omega-trajectories-vs-gamma} that the
    %frequency trajectory with small $\gamma_{30}$ tends to stay away
    %from the safe frequency bound, at the cost of having a large
    %control input, as shown in
    %plot~\subref{IEEE39-input-trajectories-vs-gamma}. We also see that a
    %large $\gamma_{30}$ may cause the controller be sensitive to
    %$\omega_{30}$, making the control input change rapidly around 8s.
  }\label{fig:trajectories-vs-gamma}
\vspace*{-3ex}
\end{figure}

\begin{figure}[tbh!]
  \centering
  \subfigure[\label{freuency-response-with-delayed-control-generator}]{\includegraphics[width=.45\linewidth]{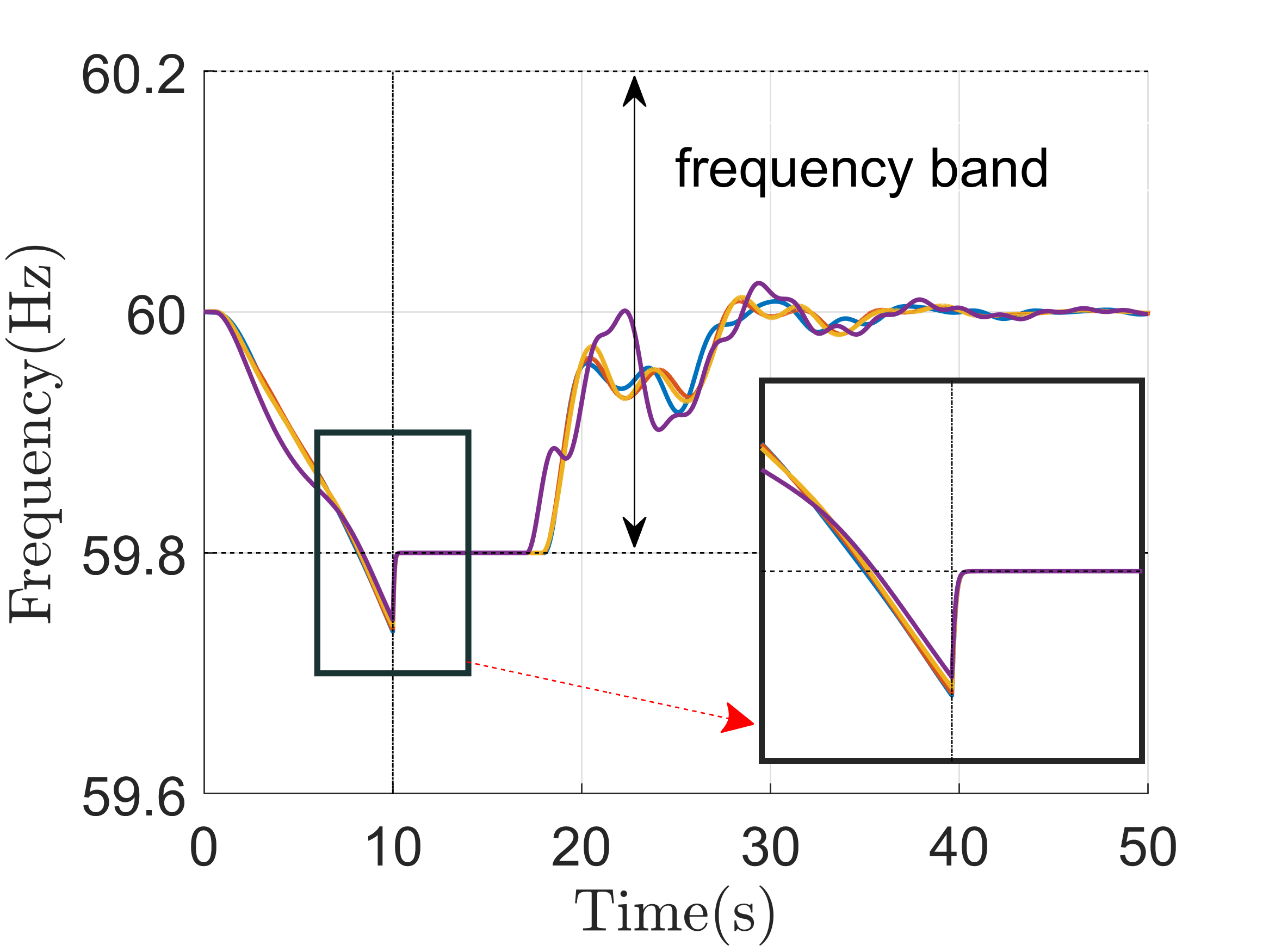}}
  \subfigure[\label{delayed-input-trajectories}]{\includegraphics[width=.45\linewidth]{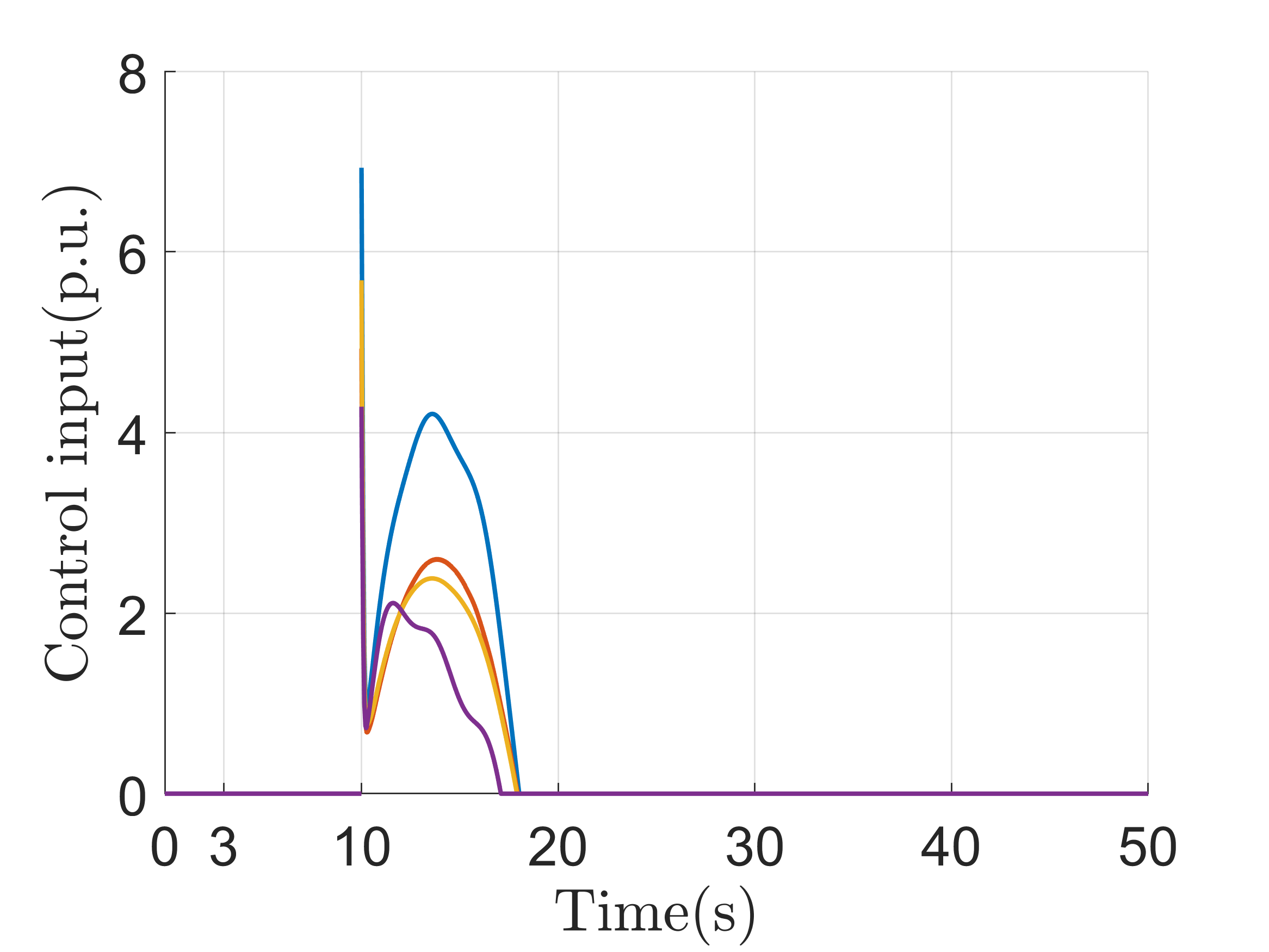}}
    \vspace{-0.3cm}
  \caption{Frequency and control input trajectories with transient
    controller available after  $t=10$s.
    % Plot~\subref{freuency-response-with-delayed-control-generator}
    %shows the frequency trajectories of the 4 generators. Due to the
    %disturbance, and without the transient controller, all 4 frequency
    %trajectories exceed the 59.8Hz safe bound at $t=10$s. However, as
    %we implement the transient controller starting from $t=10$s, those
    %unsafe trajectories come back to the safe region and never leave
    %afterwards. Plot~\subref{delayed-input-trajectories} shows the
    %control input trajectories.
    }\label{fig:delayed-trajectories}
\vspace*{-2ex}
\end{figure}

Next, we examine the effect of the choice of class-$\mathcal{K}$
function on the behavior of the transient frequency.  We focus our
attention on bus $30$ and simulate the network behavior for two
extreme values, $0.01$ and $100$, of the tunable
parameter~$\gamma_{30}$.  Figure~\ref{fig:trajectories-vs-gamma} shows
the corresponding frequency and control input trajectories for the
first 20 seconds at node 30. From
Figure~\ref{fig:trajectories-vs-gamma}\subref{IEEE39-omega-trajectories-vs-gamma},
one can see that the frequency trajectory with $\gamma_{30}=0.01$
tends to stay away from the lower safe bound (overprotection),
compared with the trajectory with large $\gamma_{30}=100$, which
results in a larger control input, as shown in
Figure~\ref{fig:trajectories-vs-gamma}\subref{IEEE39-input-trajectories-vs-gamma}.
Also, the control input with $\gamma_{30}=0.01$ is triggered earlier
around $6$s.  However, choosing a large $\gamma_{30}$ may lead to high
sensitivity.  We observe this in
Figure~\ref{fig:trajectories-vs-gamma}\subref{IEEE39-input-trajectories-vs-gamma},
as the input trajectory with $\gamma_{30}=100$ grows faster around
$8$s, compared to that with $\gamma_{30}=0.01$ does.  These
simulations verify our observations in
Remark~\ref{rmk:linear-class-K}.
Lastly, we simulate the case where some of the generator frequencies
are initially outside the safe frequency region to show how the
transient controller brings the frequencies back to the safe
region. To do so, we use the same setup as in
Figure~\ref{fig:trajectories}, but we only turn on the distributed
controller
after~$t=10s$. Figure~\ref{fig:delayed-trajectories}\subref{freuency-response-with-delayed-control-generator}
shows the frequency trajectories of the 4 generators. As the
controller is disabled for the first $10$s, all 4 frequency trajectories
are lower than 59.8Hz at $t=10$s.  After $t=10$s, all of them return to
the safe region in a monotonic way, and once they are in the region,
they never leave, in accordance with
Theorem~\ref{thm:decentralized-controller}(v).
Figure~\ref{fig:delayed-trajectories}\subref{delayed-input-trajectories}
shows the corresponding control input trajectories.

\section{Conclusions}
We have proposed a distributed transient frequency controller on a
power network that is able to maintain the nodal frequency of the
actuated buses within a given safe frequency region, and to recover it
from undesired initial conditions. We have proven that the control
input vanishes in finite time, so that the closed-loop system
possesses the same equilibrium and asymptotic convergence guarantees
as the open-loop one.  Future work will investigate the extension to
nonlinear power flow models with mechanical dynamics on generators,
the incorporation of economic cost and actuator constraints in the
controller design,
% taking advantage of the trade-offs in the choice of
% class-$\mathcal{K}$ functions for controller design,
the amelioration of control effort by having controlled nodes have
access to information beyond their neighbors, and understanding of the
connection between actuation effort and network connectivity.

\bibliographystyle{IEEEtran}
\bibliography{alias,JC,Main,Main-add}

% Generated by IEEEtran.bst, version: 1.13 (2008/09/30)
\begin{thebibliography}{10}
\providecommand{\url}[1]{#1}
\csname url@samestyle\endcsname
\providecommand{\newblock}{\relax}
\providecommand{\bibinfo}[2]{#2}
\providecommand{\BIBentrySTDinterwordspacing}{\spaceskip=0pt\relax}
\providecommand{\BIBentryALTinterwordstretchfactor}{4}
\providecommand{\BIBentryALTinterwordspacing}{\spaceskip=\fontdimen2\font plus
\BIBentryALTinterwordstretchfactor\fontdimen3\font minus
  \fontdimen4\font\relax}
\providecommand{\BIBforeignlanguage}[2]{{%
\expandafter\ifx\csname l@#1\endcsname\relax
\typeout{** WARNING: IEEEtran.bst: No hyphenation pattern has been}%
\typeout{** loaded for the language `#1'. Using the pattern for}%
\typeout{** the default language instead.}%
\else
\language=\csname l@#1\endcsname
\fi
#2}}
\providecommand{\BIBdecl}{\relax}
\BIBdecl

\bibitem{PK-JP:04}
P.~Kundur, J.~Paserba, V.~Ajjarapu, G.~Andersson, A.~Bose, C.~Canizares,
  N.~Hatziargyriou, D.~Hill, A.~Stankovic, C.~Taylor, T.~V. Cutsem, and
  V.~Vittal, ``Definition and classification of power system stability,''
  \emph{IEEE Transactions on Power Systems}, vol.~19, no.~2, pp. 1387--1401,
  2004.

\bibitem{HDC:11}
H.~D. Chiang, \emph{Direct Methods for Stability Analysis of Electric Power
  Systems: Theoretical Foundation, BCU Methodologies, and Applications}.\hskip
  1em plus 0.5em minus 0.4em\relax John Wiley and Sons, 2011.

\bibitem{FD-MC-FB:13}
F.~D{\"o}rfler, M.~Chertkov, and F.~Bullo, ``Synchronization in complex
  oscillator networks and smart grids,'' \emph{Proceedings of the National
  Academy of Sciences}, vol. 110, no.~6, pp. 2005--2010, 2013.

\bibitem{PJM-JH-JK-HJS:14}
P.~J. Menck, J.~Heitzig, J.~Kurths, and H.~J. Schellnhuber, ``How dead ends
  undermine power grid stability,'' \emph{Nature Communications}, vol.~5, no.
  3969, pp. 1--8, 2014.

\bibitem{PK:94}
P.~Kundur, \emph{Power System Stability and Control}.\hskip 1em plus 0.5em
  minus 0.4em\relax McGraw-Hill, 1994.

\bibitem{AA-EBM:06}
A.~Alam and E.~Makram, ``Transient stability constrained optimal power flow,''
  in \emph{IEEE Power and Energy Society General Meeting}, Montreal, Canada,
  Jun. 2006, electronic proceedings.

\bibitem{TTN-VLN-AK:11}
T.~T. Nguyen, V.~L. Nguyen, and A.~Karimishad, ``Transient
  stability-constrained optimal power flow for online dispatch and nodal price
  evaluation in power systems with flexible ac transmission system devices,''
  \emph{IET Generation, Transmission \& Distribution}, vol.~5, pp. 332--346,
  2011.

\bibitem{MAM-HRP-MA-MJH:14}
M.~A. Mahmud, H.~R. Pota, M.~Aldeen, and M.~J. Hossain, ``Partial feedback
  linearizing excitation controller for multimachine power systems to improve
  transient stability,'' \emph{IEEE Transactions on Power Systems}, vol.~29,
  pp. 561--571, 2014.

\bibitem{TSB-TL-DJH:15}
T.~S. Borsche, T.~Liu, and D.~J. Hill, ``Effects of rotational inertia on power
  system damping and frequency transients,'' in \emph{{IEEE} Conf.\ on Decision
  and Control}, Osaka, Japan, 2015, pp. 5940--5946.

\bibitem{BKP-SB-FD:17}
B.~K. Poolla, S.~Bolognani, and F.~Dorfler, ``Optimal placement of virtual
  inertia in power grids,'' \emph{IEEE Transactions on Automatic Control},
  2017, to appear.

\bibitem{MA:14}
M.~Althoff, ``Formal and compositional analysis of power systems using
  reachable sets,'' \emph{IEEE Transactions on Power Systems}, vol.~29, no.~5,
  pp. 2270--2280, 2014.

\bibitem{YCC-ADD:12}
Y.~C. Chen and A.~D. Dom{\'i}nguez-Garc{\'i}a, ``A method to study the effect
  of renewable resource variability on power system dynamics,'' \emph{IEEE
  Transactions on Power Systems}, vol.~27, no.~4, pp. 1978--1989, 2012.

\bibitem{YZ-JC:17-acc}
Y.~Zhang and J.~Cort\'es, ``Transient-state feasibility set approximation of
  power networks against disturbances of unknown amplitude,'' in
  \emph{{A}merican {C}ontrol {C}onference}, Seattle, WA, May 2017, pp.
  2767--2772.

\bibitem{ADA-XX-JWG-PT:16}
A.~D. Ames, X.~Xu, J.~W. Grizzle, and P.~Tabuada, ``Control barrier function
  based quadratic programs for safety critical systems,'' \emph{IEEE
  Transactions on Automatic Control}, vol.~62, no.~8, pp. 3861--3876, 2017.

\bibitem{FB-SM:08}
F.~Blancini and S.~Miani, \emph{Set-theoretic Methods in Control}.\hskip 1em
  plus 0.5em minus 0.4em\relax Boston, MA: Birkh{\"a}user, 2008.

\bibitem{FB-JC-SM:08cor}
F.~Bullo, J.~Cort{\'e}s, and S.~Mart{\'\i}nez, \emph{Distributed Control of
  Robotic Networks}, ser. Applied Mathematics Series.\hskip 1em plus 0.5em
  minus 0.4em\relax Princeton University Press, 2009, electronically available
  at \url{http://coordinationbook.info}.

\bibitem{NB:94}
N.~Biggs, \emph{Algebraic Graph Theory}, 2nd~ed.\hskip 1em plus 0.5em minus
  0.4em\relax Cambridge University Press, 1994.

\bibitem{CZ-UT-NL-SL:14}
C.~Zhao, U.~Topcu, N.~Li, and S.~H. Low, ``Design and stability of load-side
  primary frequency control in power systems,'' \emph{IEEE Transactions on
  Automatic Control}, vol.~59, no.~5, pp. 1177--1189, 2014.

\bibitem{EM:14}
E.~Mallada, ``Distributed network synchronization: The internet and electric
  power grids,'' Ph.D. dissertation, Cornell University, 2014, electronically
  available at
  \url{http://www.its.caltech.edu/\~{}mallada/pubs/2014/thesis.pdf}.

\bibitem{JM-JWB-JRB:08}
J.~Machowski, J.~W. Bialek, and J.~R. Bumby, \emph{Power System Dynamics:
  Stability and Control}.\hskip 1em plus 0.5em minus 0.4em\relax Chichester,
  England: Wiley, 2008.

\bibitem{KWC-JC-GR:09}
K.~W. Cheung, J.~Chow, and G.~Rogers, \emph{Power System Toolbox, v 3.0.}\hskip
  1em plus 0.5em minus 0.4em\relax Rensselaer Polytechnic Institute and Cherry
  Tree Scientific Software, 2009.

\end{thebibliography}

\end{document}